\documentclass[twocolumn,aps,superscriptaddress,showpacs,floatfix]{revtex4-1}
\usepackage{graphicx}
\usepackage{dcolumn}
\usepackage{bm}
\usepackage[colorlinks,linkcolor=blue,anchorcolor=blue,citecolor=blue,urlcolor=blue]{hyperref}
\usepackage{multirow}
\usepackage{textcomp}
\usepackage{color}
\usepackage{ulem}
\usepackage{soul}
\usepackage{subfigure}
\usepackage{array}
\usepackage{makecell}
\usepackage{verbatim}
\usepackage{threeparttable}
\usepackage{float}
\usepackage{appendix}
\usepackage{amsmath}
\usepackage{cancel}
\begin{document}
\preprint{APS/123-QED}

\title{New measurement of the elemental fragmentation cross sections of 218~MeV/nucleon $^{28}$Si on a carbon target}

\author{Guang-Shuai Li}
\affiliation{School of Physics, Beihang University,  Beijing 100191, People's Republic of China}

\author{Jun Su}
\affiliation{Sino-French Institute of Nuclear Engineering and Technology, Sun Yat-sen University, Zhuhai 519082, China}

\author{Bao-Hua Sun}
\email{Corresponding author: bhsun@buaa.edu.cn}
\affiliation{School of Physics, Beihang University,  Beijing 100191, People's Republic of China}

\author{Satoru Terashima}
\affiliation{School of Physics, Beihang University,  Beijing 100191, People's Republic of China}
\affiliation{Institute of Modern Physics, Chinese Academy of Sciences, Lanzhou 730000, China}

\author{Jian-Wei Zhao}
\email{Corresponding author: j.zhao@pku.edu.cn}
\affiliation{School of Physics, Beihang University,  Beijing 100191, People's Republic of China}
\affiliation{State Key Laboratory of Nuclear Physics and Technology, School of Physics, Peking University, Beijing 100871, China}

\author{Xiao-Dong Xu}
\affiliation{Institute of Modern Physics, Chinese Academy of Sciences, Lanzhou 730000, China}

\author{Ji-Chao Zhang}
\affiliation{School of Physics, Beihang University,  Beijing 100191, People's Republic of China}

\author{Ge Guo}
\affiliation{School of Physics, Beihang University,  Beijing 100191, People's Republic of China}

\author{Liu-Chun He}
\affiliation{School of Physics, Beihang University,  Beijing 100191, People's Republic of China}

\author{Wei-Ping Lin}
\affiliation{Key Laboratory of Radiation Physics and Technology of the Ministry of Education, Institute of Nuclear Science and
Technology, Sichuan University, Chengdu 610064, China}

\author{Wen-Jian Lin}
\affiliation{School of Physics, Beihang University,  Beijing 100191, People's Republic of China}

\author{Chuan-Ye Liu}
\affiliation{School of Physics, Beihang University,  Beijing 100191, People's Republic of China}

\author{Chen-Gui Lu}
\affiliation{Institute of Modern Physics, Chinese Academy of Sciences, Lanzhou 730000, China}

\author{Bo Mei}
\affiliation{Institute of Modern Physics, Chinese Academy of Sciences, Lanzhou 730000, China}
\affiliation{Sino-French Institute of Nuclear Engineering and Technology, Sun Yat-sen University, Zhuhai 519082, China}

\author{Zhi-Yu Sun}
\affiliation{Institute of Modern Physics, Chinese Academy of Sciences, Lanzhou 730000, China}

\author{Isao Tanihata}
\affiliation{School of Physics, Beihang University,  Beijing 100191, People's Republic of China}

\author{Meng Wang}
\affiliation{School of Physics, Beihang University,  Beijing 100191, People's Republic of China}

\author{Feng Wang}
\affiliation{School of Physics, Beihang University,  Beijing 100191, People's Republic of China}

\author{Shi-Tao Wang}
\affiliation{Institute of Modern Physics, Chinese Academy of Sciences, Lanzhou 730000, China}

\author{Xiu-Lin Wei}
\affiliation{School of Physics, Beihang University,  Beijing 100191, People's Republic of China}

\author{Jing Wang}
\affiliation{School of Physics, Beihang University,  Beijing 100191, People's Republic of China}

\author{Jun-Yao Xu}
\affiliation{School of Physics, Beihang University,  Beijing 100191, People's Republic of China}

\author{Jin-Rong Liu}
\affiliation{School of Physics, Beihang University,  Beijing 100191, People's Republic of China}

\author{Mei-Xue Zhang}
\affiliation{School of Physics, Beihang University,  Beijing 100191, People's Republic of China}

\author{Yong Zheng}
\affiliation{Institute of Modern Physics, Chinese Academy of Sciences, Lanzhou 730000, China}

\author{Li-Hua Zhu}
\affiliation{School of Physics, Beihang University,  Beijing 100191, People's Republic of China}

\author{Xue-Heng Zhang}
\affiliation{Institute of Modern Physics, Chinese Academy of Sciences, Lanzhou 730000, China}

\date{\today}

\begin{abstract}
Elemental fragmentation cross sections (EFCSs) of stable and unstable nuclides have been investigated with various projectile-target combinations at a wide range of incident energies. These data are critical to constrain and develop the theoretical reaction models and to study the propagation of galactic cosmic rays (GCR).
In this work, 
we present a new EFCS measurement for $^{28}$Si on carbon at 218~MeV/nucleon performed at the Heavy Ion Research Facility (HIRFL-CSR) complex in Lanzhou. The impact of the target thickness has been well corrected to derive an accurate EFCS.
Our present results with charge changes $\Delta Z$ = 1-6 are compared to the previous measurements and to the predictions from the models modified EPAX2, EPAX3, FRACS, ABRABLA07, NUCFRG2, and IQMD coupled with GEMINI (IQMD+GEMINI).
All the models fail to describe the odd-even staggering strength in the elemental distribution, with the exception of the IQMD+GEMINI model, 
which can reproduce the EFCSs with an accuracy of better than 3.5\% for $\Delta Z\leq5$.
The IQMD+GEMINI analysis shows that the odd-even staggering in EFCSs occurs in the sequential statistical decay stage rather than in the initial dynamical collision stage. This offers a reasonable approach to understand the underlying mechanism of fragmentation reactions.

\end{abstract}
\maketitle

\section{Introduction}
\label{section1}

One of the experimental focuses in current nuclear physics is to determine the fragmentation cross sections of nuclei with different isospins and masses~\cite{MA2021103911}. The nuclei of interest can be produced via projectile fragmentation. 
In the past several decades, systematic investigations in experiments have provided a valuable fragmentation cross section database. 
This is crucial to understand the interaction of energetic heavy nuclei.
Moreover, the cross section data involving stable nuclei are essential to understanding 
the shielding and the radiation protection of energetic heavy ions.

Silicon is one of the “primary” cosmic rays, and 
known to be the eighth most common element in the universe by mass. It is most widely distributed  in cosmic dust, planetoids, and planets as various forms of silicon dioxide (silica) or silicates.
The relevant fragmentation or spallation cross section data on hydrogen, helium, or carbon at relativistic energies are important inputs to investigate the propagation of galactic cosmic rays (GCR) (see, e.g., Ref.~\cite{RevModPhys.83.1245}).
Moreover, silicon is the main integrated-circuit element of onboard spacecraft electronics.
Its interactions with high-energy charged particles in the cosmic rays are the underlying reason for malfunctions in integrated circuits.

Over the last two decades, experimental studies on the fragmentation of $^{28}$Si have been performed at various energies by using different detection techniques in laboratories worldwide~\cite{PhysRevC.41.533,FLESCH2001237,ZEITLIN2007341,CECCHINI2008206,SAWAHATA2017142,LI2016314,Li_2017}.
This isotope offers an ideal benchmark dataset to check the consistency of experiments, to evaluate the model reliability, and further to investigate the nuclear structure underneath the heavy-ion collisions. 
Nevertheless, we are aware that although experimental data for $^{28}$Si are relatively rich, different energy-dependent behaviors were reported.
Moreover, the elemental fragmentation cross sections (EFCSs) performed at similar incident energies for the same reaction system~\cite{PhysRevC.41.533,ZEITLIN2007341,SAWAHATA2017142} present a large discrepancy by up to about 40\%, albeit with similar heavy-ion detector systems.
In addition, the cross sections at 700-800 MeV/nucleon~\cite{LI2016314,Li_2017}, which were extracted by using CR-39 detectors, have relatively large uncertainties and are systematically larger than those measured by the active heavy-ion detectors. 
Such divergence in experimental data could be due to the systematic uncertainties in the relevant experiments. 
This hampers our understanding of the underlying reaction mechanism in fragmentation reactions, e.g., the formation of odd-even staggering in charge distribution~\cite{PhysRevC.83.014608,Cheng_2012,Cheng_2015}.

Several models are now available to predict the heavy-ion reaction cross sections at intermediate energies. 
Empirical parametrizations, such as EPAX3~\cite{PhysRevC.86.014601}, modified EPAX2~\cite{ZHANG201359}, and FRACS~\cite{PhysRevC.95.034608} models, are developed relying on optimizing the parameters to the existing experimental data.
In the EPAX3 empirical formula, 
cross sections are considered to be independent of incident projectile energies, 
while the modified EPAX2 model includes
the target isospin and incident projectile energy dependence for the cross sections at intermediate energies of 20-200 MeV/nucleon.
In the FRACS model,
the target dependence, incident projectile energy as well as odd-even staggering of cross sections are taken into account.

In addition to the empirical formulas, statistical models (such as ABRABLA07 and NUCFRG2) have been developed. 
In this framework, highly excited prefragments are formed first, and decay subsequently by emitting light particles ($p$, $d$, $t$, $\alpha$, etc.),
until the resulting products are unable to undergo further decay. 
The NUCFRG2 model~\cite{WILSON199495} is a semiempirical cross section model based on the abrasion-ablation formulation~\cite{PhysRevC.12.1888}.

The isospin-independent quantum molecular dynamics model (IQMD)~\cite{HARTNACK1989303} 
coupled with the GEMINI model~\cite{CHARITY1988371} (hereafter referred to as IQMD+GEMINI) is a hybrid model, 
in which the IQMD model and GEMINI model are applied separately to describe the primary violent state of heavy-ion collisions to form the hot prefragments, and the subsequent
statistical deexcitation of prefragments. 

Since 2015, we have performed a new experimental campaign in the energy range of 200-500 MeV/nucleon, aiming to determine the EFCSs of more than ten stable isotopes. Here we report the first result of $^{28}$Si on carbon at 218 MeV/nucleon.
The experiment overview is described in Sec.~\ref{section2}, and the data analysis procedure is detailed in Sec.~\ref{section3}.
Then we analyze the experimental data by comparing with previous measurements and model predictions,
and discuss the odd-even staggering in fragment cross section in Sec.~\ref{section4}. 
Finally, a summary is given in Sec.~\ref{section5}. 

\section{EXPERIMENT}
\label{section2}

The experiment was carried out at the Heavy Ion Research Facility (HIRFL-CSR) in Lanzhou~\cite{XIA200211,ZHAN2010694c}. $^{40}$Ar primary beam was accelerated to 320 MeV/nucleon in the heavy-ion synchrotron CSRm, extracted in the slow extraction mode, and impinged into a 10 mm thick beryllium production target. 
The target was placed at the entrance of the Second Radioactive Ion Beam Line in Lanzhou (RIBLL2)~\cite{SUN201878}.
The secondary $^{28}$Si beam produced via projectile fragmentation was 
then separated in flight by setting the first half of RIBLL2 to a magnetic rigidity ($\textit{B}\rho$) of 5.03 Tm, 
and transported to the External Target Facility (ETF).
There, the $^{28}$Si beam was directed into a natural carbon target with the thickness of 1.86 g/$\rm{cm^{2}}$.

\begin{figure}[htbp!]
\centering
\includegraphics[width=0.46\textwidth]{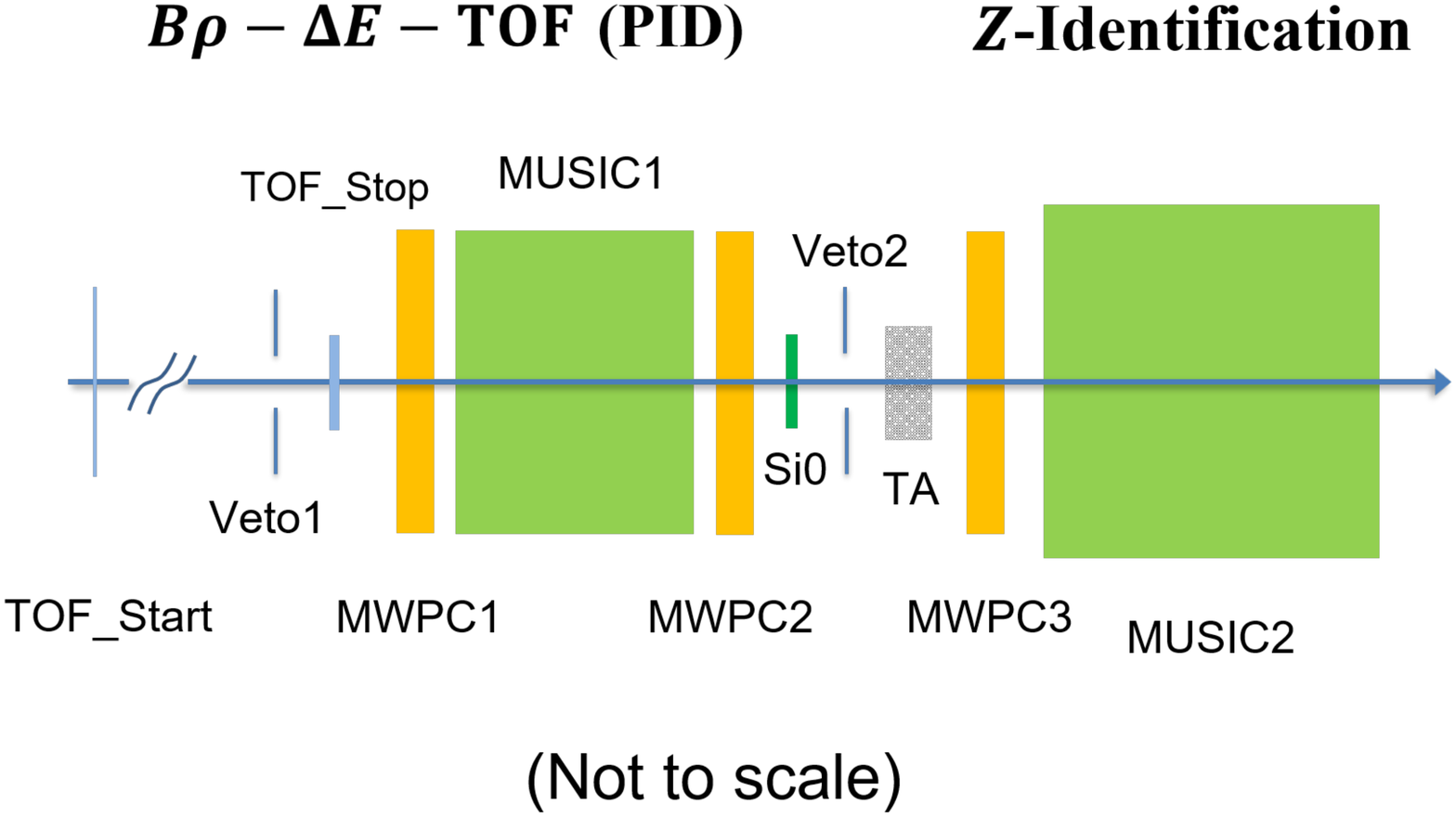}
\caption{(Color online) Schematic layout of the experimental detector setup at F1 and External Target Facility (ETF). For details, see the text.} 
\label{fig1}
\end{figure}

A schematic layout of the experimental detector setup is illustrated in Fig.~\ref{fig1}. The time of flight (TOF) of $^{28}$Si beam from the first dispersive focal plane (F1) of RIBLL2 to ETF was determined event by event by a pair of fast timing plastic scintillators placed 
at F1 and ETF, respectively.
The corresponding flight length is approximately 26 m.
The intrinsic TOF resolution achieved from detectors is as low as 80 ps ($\sigma$)~\cite{SUN201878,Lin_2017,ZHAO201995,Zhao:2020seq}.

A pair of multiple sampling ionization chambers~\cite{ZHANG2015389,SUN2019390} (denoted as MUSIC1 and MUSIC2 in Fig.~\ref{fig1}) was installed 
at both sides of
the carbon reaction target (defined as TA in Fig.~\ref{fig1}) to measure the energy deposition of the incident and outgoing particles, respectively.
The respective active areas are 85 mm $\times$ 85~mm and 130~mm $\times$ 130~mm.
The large acceptance of MUSIC2 guarantees the full coverage of outgoing particles.
The total length of 630 mm of MUSIC2 is divided into 24 cells by 13 anodes and 12 cathodes.
In total, six signals were readout by connecting the neighboring four cells in series to one charge-sensitive preamplifier.
Two  types of preamplifiers have been
used for the first five subchambers and the last one in this experiment, respectively.
A $\textit{Z}$ resolution of 0.25-0.35 (FWHM) was achieved for fragments of $^{40}$Ar using the first five subchambers~\cite{ZHAO201995}.

Two 100 mm $\times$ 100 mm plastic scintillators, each with a 30 mm~$\times$~30 mm hole (denoted as Veto1 and Veto2 in Fig.~\ref{fig1}) at the center, were positioned
upstream of the carbon reaction target to 
limit the momentum dispersion of incoming beams, and to reduce the total trigger rate.
Three multiwire proportional chambers (denoted as MWPCs in Fig.~\ref{fig1}) with an effective area of 85 mm $\times$ 85 mm, 
two upstream and one downstream of the reaction target, were mounted to monitor the trajectories of incoming and outgoing particles.
Combining the information can help to select the ``best" incident beams within a certain angle and position acceptance. 
In the data analysis hereafter, we have restricted the beam size to 25 mm $\times$ 25 mm on the target.

Measurements without the reaction target were carried out to eliminate the effect of reactions in the materials other than the target (e.g., detectors).
For the case of an infinitely thin target, the attenuation effect of incident beam and the secondary reactions in the target can be safely neglected.
This then results in the commonly used cross section formula, i.e., the EFCS from incident beam with $Z_{\rm{i}}$ to an element with $Z$ follows 
\begin{equation}
\begin{aligned}
\sigma_{\Delta Z=Z_{\rm{i}}-Z}=
\frac{1}{t}(\frac{N_{\rm{F}}}{N_{0}}-\frac{N^{0}_{\rm{F}}}{N^{0}_{0}})\;,
\end{aligned}
\label{eq1}
\end{equation}
where
$N_{0}$ ($N^{0}_{0}$) and $N_{\rm{F}}$ ($N^{0}_{\rm{F}}$)
represent the counts of incident beam and fragment species with $Z$, for the target-in (target-out) case, respectively. 
$t$ denotes the number of target nuclei per unit area.
In this work, the target thickness is about one-ninth of mean free path ($\approx$ 90 mm) of $^{28}$Si on carbon target.
The attenuation effect of incident $^{28}$Si and the secondary reactions in this case could have a considerable impact.
To account for these effects, we have deduced the formula of EFCS as detailed in the Appendix.
The EFCS and its uncertainty can be evaluated by considering the two extreme cases: 
one neglecting the contributions to the desired fragments from secondary reactions, and the other neglecting the contributions from the cascade fragmentation processes.
The solution of the two cases corresponds to Eqs.~(\ref{eqA4}) and (\ref{eqA12}). 
The final centroid values of EFCS are determined by Eq.~(\ref{eqA4}) for charge changes $\Delta Z$ = 1-2,  
by averaging the cross sections from Eqs.~(\ref{eqA4}) and (\ref{eqA12}) for $\Delta Z$ = 3-4, and 
by Eq.~(\ref{eqA12}) for $\Delta Z$ = 5-6, respectively.
The statistical errors for $\Delta Z$ = 1-4 and 6 are 4.9-6.5\%, and 8.4\% for $\Delta Z$ = 5, respectively.
The systematic errors are estimated to be 2.7-6.0\% by considering the peak decoupling methods, the beam attenuation and secondary reactions in the target for $\Delta Z$ = 1-6. The final uncertainties are detailed in the Appendix.

\section{Data analysis}
\label{section3}

\begin{figure}[htpb!]
\centering
\includegraphics[width=0.45\textwidth]{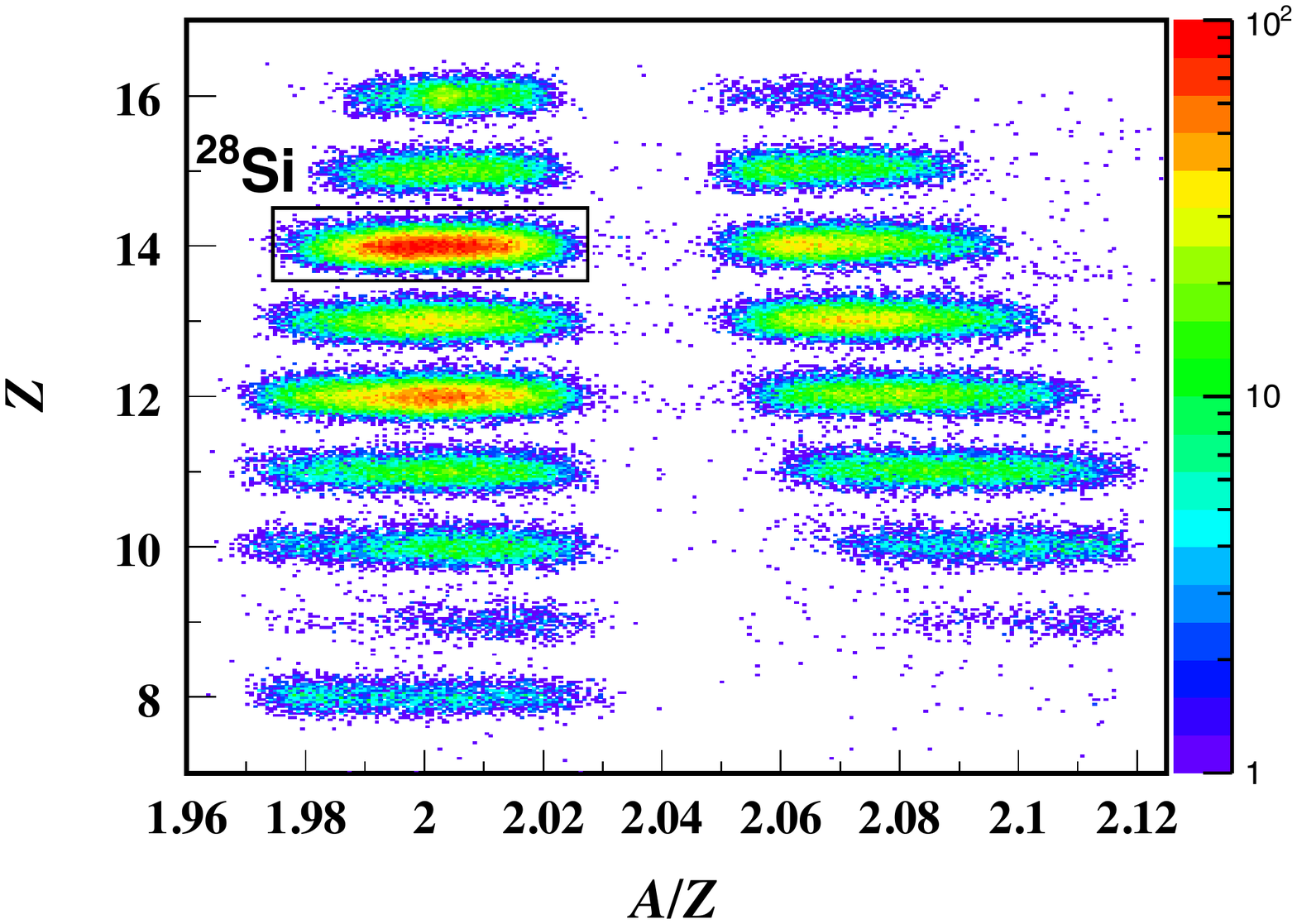}
\caption{(Color online) Particle identification of cocktail beams produced via $^{40}$Ar + $^{9}$Be at 320 MeV/nucleon. 
The $^{28}$Si beam is selected by setting the rectangular gate.}
\label{fig2}
\end{figure}

The particle identification with respect to atomic number $Z$ and mass-to-charge ratio $A$/$Z$ is presented in Fig.~\ref{fig2}, by means of the energy deposition of incident particles ($\Delta E$) and TOF measurements.
In the figure one can see that various cocktail beams can be clearly identified.
The $^{28}$Si events are selected with a high purity by setting a rectangular gate as indicated in Fig.~\ref{fig2}, which corresponds to $\pm$ 3.4$\sigma$ in $Z$ projection. 
The $^{28}$Si events are counted as $\textit{N}_{0}$. 
The contamination level from the overlap of neighboring $\textit{Z}$ = 13 and $\textit{Z}$ = 15 beam events relative to $\textit{Z}$ = 14 is evaluated to be less than 10$^{-4}$. 

\begin{figure}[htpb!]
\centering
\includegraphics[width=0.45\textwidth]{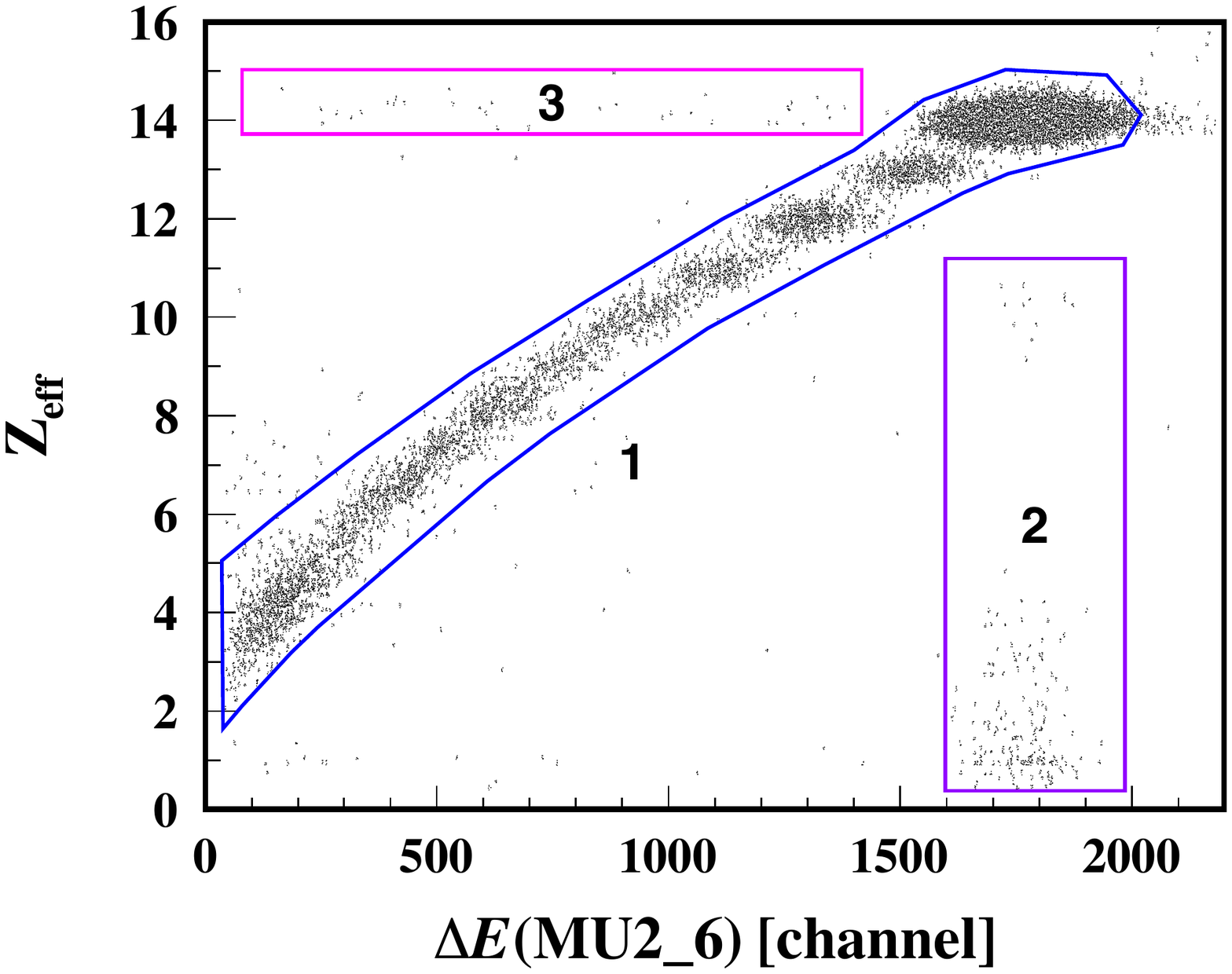}
\caption{(Color online) Effective charge ($Z_{\rm{eff}}$) identified using the first five subchambers of MUSIC2 against the energy deposition in the last subchamber of MUSIC2. Refer to the text for details.}
\label{fig3}
\end{figure}

The incident energy of $^{28}$Si in the middle of the carbon target is determined to be 218 MeV/nucleon.
The effective $Z$ distribution ($Z_{\rm{eff}}$) of fragments of $^{28}$Si versus the energy deposition in the last subchamber of MUSIC2 [denoted as $\Delta E$(MU2\_6)], is shown in Fig.~\ref{fig3}.
The $Z_{\rm{eff}}$ value is calibrated from the energy deposition in the first five subchambers of MUSIC2 from $^{16}$O to $^{32}$S with $A/Z$ = 2 produced by the $^{40}$Ar primary beam.
The calibration used is $\Delta E=C_{1}\times Z_{\rm{eff}}^{C2}+C_{3}$, and three parameters are determined separately by the best fit.
Each cluster in region 1 represents a particular fragment species. 
The events at charge $Z$ = 14 are dominant and can be easily identified.
The peak positions for particles with $Z$ = 9 to 13 are less distinct but can be fairly well distinguished, whereas it is difficult to identify and separate the isotopes with $Z$~$\leq$ 8.
Furthermore, the events in region 2
are identified as $Z$ = 14 using the last subchamber of MUSIC2, but their energy depositions are not fully registered in the first five subchambers.
This accounts for approximately 0.5\% of incident $^{28}$Si.
Events in region 3, the counterpart of region 2, 
can be identified by using the first five subchambers, and are about 0.1\% of $^{28}$Si.
The events with $Z_\text{eff}<2$ are due to the effect induced by the detection efficiency of the first five subchambers.
This accounts for approximately 0.4\% of incident $^{28}$Si. 
The events in region 1 are selected hereafter 
for the cross section determination since regions 2 and 3 would not affect the counting of lower $Z$ events.

\begin{figure}[htpb!]
\centering
\includegraphics[width=0.48\textwidth]{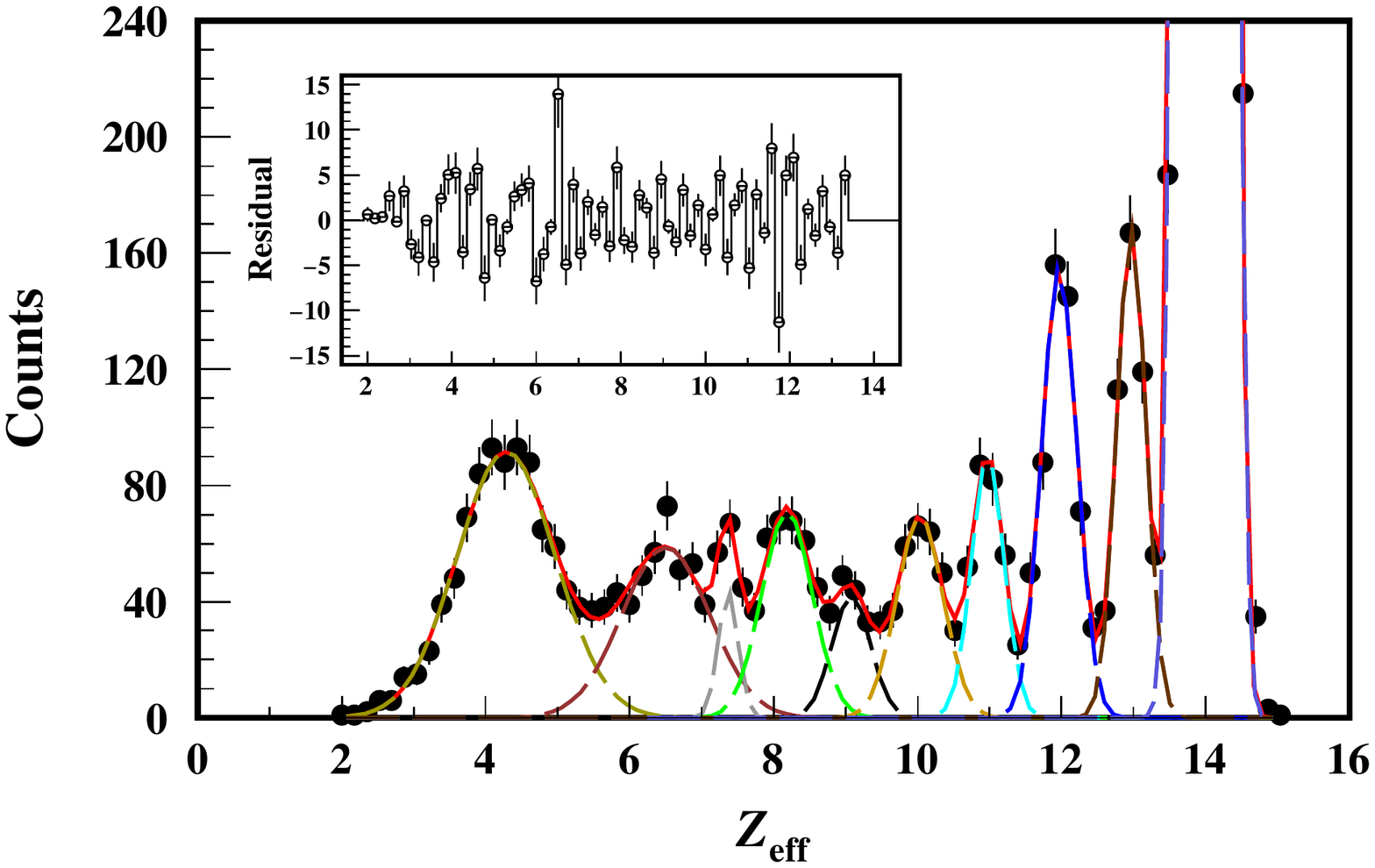}
\caption{(Color online) $Z_{\rm{eff}}$ distribution (black dots) of fragments produced by the $^{28}$Si beam impinging on a carbon target at 218 MeV/nucleon. 
A multiple-Gaussian function (red solid line) is used to describe the charge distribution with $Z$ = 2-14. 
The individual Gaussian component for a particular peak is shown by the dotted line in different colors. The residual distribution from the global fit is shown in the inset.}
\label{fig4}
\end{figure}

The one-dimensional $Z_{\rm{eff}}$ distribution 
of $^{28}$Si fragments is presented in Fig.~\ref{fig4}. 
One can see that the peak positions at $Z$ = 6 and 7 are shifted to the positions higher than their nominal values. 
The peak width shows a generally increasing trend with decreasing $Z$,   
but with a kink at $Z$ = 6 and 7, and is relatively large at $Z=$ 6, 8, and 10.
Besides, the peaks of fragments with $Z \leq 5$ are mixed with each other, thus cannot be resolved. 
We performed a cross-check on the resolution of MUSIC2 for $Z$~$\leq$ 8, and concluded that MUSIC2 is able to resolve the fragments with $Z$~$\geq$ 5.
This was done by selecting $^{16}$O as the incident beam instead.
Thus, the unresolved peaks in Fig.~\ref{fig4} are not due to the charge $Z$ resolution.

In our experiment, we employed MUSIC2 with a large geometric acceptance.
This size is essential to cover all nonreacted incident elements~\cite{Zhao:2020seq,BaoHS2020CSB} that have a similar velocity as the incident $^{28}$Si, and guarantee a good charge $Z$ resolution.
As noted in Refs.~\cite{PhysRevC.41.533,PhysRevC.64.024902,ZEITLIN2007341},
the multiplicity of light fragments will increase with decreasing atomic number of the produced leading fragment.
When the atomic number $Z$ of produced fragments is much smaller than that of the incident particle, the large acceptance detector will also cover 
the majority of
the lighter particles in addition to the leading fragments. 
This results in the shift of charge peak positions to higher values than expected for isotopes with $Z=6$ and 7, and less distinct peaks and largely overlapping beam events for isotopes with $Z$~$\leq$ 5.
Another consequence of the large acceptance detector is the broadening in peak widths, in particular when the $\alpha$ particles can be produced simultaneously with the leading fragments ($\textit{e.g.}$, with $N=Z$).
The protons from fragmentation do not play a significant role in shaping the final peak width due to their relatively small energy deposition in MUSIC2.

The $Z_{\rm{eff}}$ distribution from $Z$ = 6 to 14 can be decoupled with a multiple-Gaussian function in which the peak position and width for each $Z_{\rm{eff}}$ can be confined well from experimental data. 
Furthermore, we find that the $Z_{\rm{eff}}$ = 2-5.5 distribution can be nicely reproduced with a Gaussian function.
Hence the total beam events corresponding to $Z$ = 2-5 and $Z$ = 6-14 have been decoupled using the individual Gaussian components.
The residual distribution, which is defined as the difference between experimental data and the global fit, scatters randomly around zero as shown in the inset of Fig.~\ref{fig4}.
The overall reduced chi-square value 
is determined to be 0.60 from $Z_{\rm{eff}}$ = 2 to 13.5.

In this study, we focus on EFCSs with $Z$ = 8-13. To determine cross sections for $Z\leq7$ is beyond our scope. 
Considering the low statistics, the cross section may suffer from the peak decoupling methods, depending on the multiple-Gaussian peak widths and central values.
As a test, we have refitted the data in the same approach but with fixed peak widths
of $\Delta Z$ = 3-6 to those extrapolated linearly from $\Delta Z$ = 0-2.
It is found that the cross sections agree well with those obtained without fixed widths within the statistical uncertainties, and the centroid values increase by 6.5 and 5.6 mb for $\Delta Z$ = 3 and 5, and decrease by 11.0 and 9.6 mb for $\Delta Z$ = 4 and 6 relative to those
determined without fixed peak widths, respectively.
As a result, we have treated the potential uncertainty in fitting approach, beam attenuation, and secondary reactions in the target as a systematic uncertainty. 
The final cross sections for $\Delta Z$ = 1-6 and the relevant uncertainties are given in Table~\ref{tab1}.

\begin{table*}[htpb!]
    \centering
    \caption{Summary of the elemental fragmentation and total charge-changing cross section data for $^{28}$Si + $^{12}$C at incident energies ranging from 200 to 800 MeV/nucleon. The first and second parentheses in our present data are the statistical and systematic uncertainties, respectively.}
    \setlength{\tabcolsep}{0.4mm}{
    \begin{tabular}{ccccccccccccc}
    \hline
    \hline
        \multirow{2}{*}{Isotope}
        &Incident energy
        &$\sigma_{\Delta\textit{Z}=1}$
        &$\sigma_{\Delta\textit{Z}=2}$   
        &$\sigma_{\Delta\textit{Z}=3}$
        &$\sigma_{\Delta\textit{Z}=4}$    
        &$\sigma_{\Delta\textit{Z}=5}$    
        &$\sigma_{\Delta\textit{Z}=6}$    
        &$\sigma_{\Delta\textit{Z}=7}$
        &$\sigma_{\Delta\textit{Z}=8}$
        &$\sigma_{\Delta\textit{Z}=9}$        
        &$\sigma_{\rm{CC}}$
        &\multirow{2}{*}{Reference}\\
& MeV/nucleon & mb & mb & mb & mb & mb & mb & mb & mb & mb &mb\\
    \hline
\multirow{12}{*}{$^{28}$Si}
&\multirow{1}{*}{218} &140(7)(4) &155(8)(4) &90(6)(4) &101(6)(6) &51(4)(3) &114(6)(5) & & &
&1126(21) &Present work\\

&266 & 140(7) &164(8) &92(4) &94(5) &51(3) &94(8) &73(6) &94(8) & &1131(34)\footnote{Incident energy: 262 to 278 MeV/nucleon.} &~\cite{ZEITLIN2007341}\\

&268 &123(2) &139(2) &75(1) &84(1) &46(1) &92(1) &62(1) &103(1) &
&1106(5) &~\cite{SAWAHATA2017142}\\

&344 &122(4) &143(4) &80(3) &86(3) &45(2) &93(6) &71(5) &104(7) & &1125(16)\footnote{Incident energy: 340 to 364 MeV/nucleon.} &~\cite{ZEITLIN2007341}\\

&467 &125(5) &130(5) &67(3) &77(4) &38(2) &79(4) &62(3) &85(4)  & &1136(13) &~\cite{FLESCH2001237}\\

&503 &130(2) &141(2) &68(2) &72(2) &31(1) &68(2) &40(2) &73(4) & &1176(12)\footnote{From Ref.~\cite{PhysRevC.41.520}\label{webber}.} &~\cite{PhysRevC.41.533}\\

&560 &118(3) &134(3) &74(2) &80(2) &42(1) &86(5) &65(4) &98(6) & &1142(16)\footnote{Incident energy: 536 to 568 MeV/nucleon.} &~\cite{ZEITLIN2007341}\\

&723 &205(20)&131(16)&58(11)&75(12)&33(8) &58(11)&62(11)&81(13)& &1186(42)                                                            &~\cite{Li_2017}\\

&736 &254(24) &131(17) &65(12) &100(15) &65(12) &100(15) &89(14) &104(16)
& 21(7) &1179(50) &~\cite{Li_2017}\\

&765 &113(3) &124(3) &68(2) &73(2) &41(1) &80(5) &64(4) &91(5) & &1110(14)\footnote{Incident energy: 760 to 770 MeV/nucleon.} &~\cite{ZEITLIN2007341}\\

&770 &157(2) &163(2) &64(2) &63(2) &30(2) &65(2) &48(1) &74(2) & &1183(12)\textsuperscript{\ref{webber}} &~\cite{PhysRevC.41.533}\\

&788 &225(19) &154(15) &92(12) &75(11) &49(9) &92(12) &55(9)
&107(13) &50(9) &1127(42) &~\cite{LI2016314}\\
   \hline
   \hline
    \end{tabular}}
    \label{tab1}
\end{table*}

\section{RESULTS AND DISCUSSIONS}
\label{section4}
\subsection{Elemental fragmentation cross section}

Table~\ref{tab1} summarizes the EFCSs obtained in this work along with previous measurements for reactions of $^{28}$Si + $^{12}$C at incident energies from 200 to 800 MeV/nucleon. The first and second parentheses in our present data are the statistical and systematic uncertainties, respectively.
The experimental data can be classified into two groups by using either active detectors (like silicon or ionization chambers)~\cite{PhysRevC.41.533,ZEITLIN2007341,SAWAHATA2017142} or CR-39 detectors~\cite{FLESCH2001237,LI2016314,Li_2017}.
Our present measurements for charge changes $\Delta Z$ = 1-6 at 218~MeV/nucleon are generally consistent with those at 266 and 268~MeV/nucleon within uncertainties.
Meanwhile, the total charge-changing cross section, $\sigma_{\rm{CC}}$, is extracted to be 1126(21)~mb.
The charge-changing cross sections appear generally constant above 200~MeV/nucleon~\cite{PhysRevC.82.014609}.

\begin{figure*}
    \centering
    \includegraphics[width=0.87\textwidth]{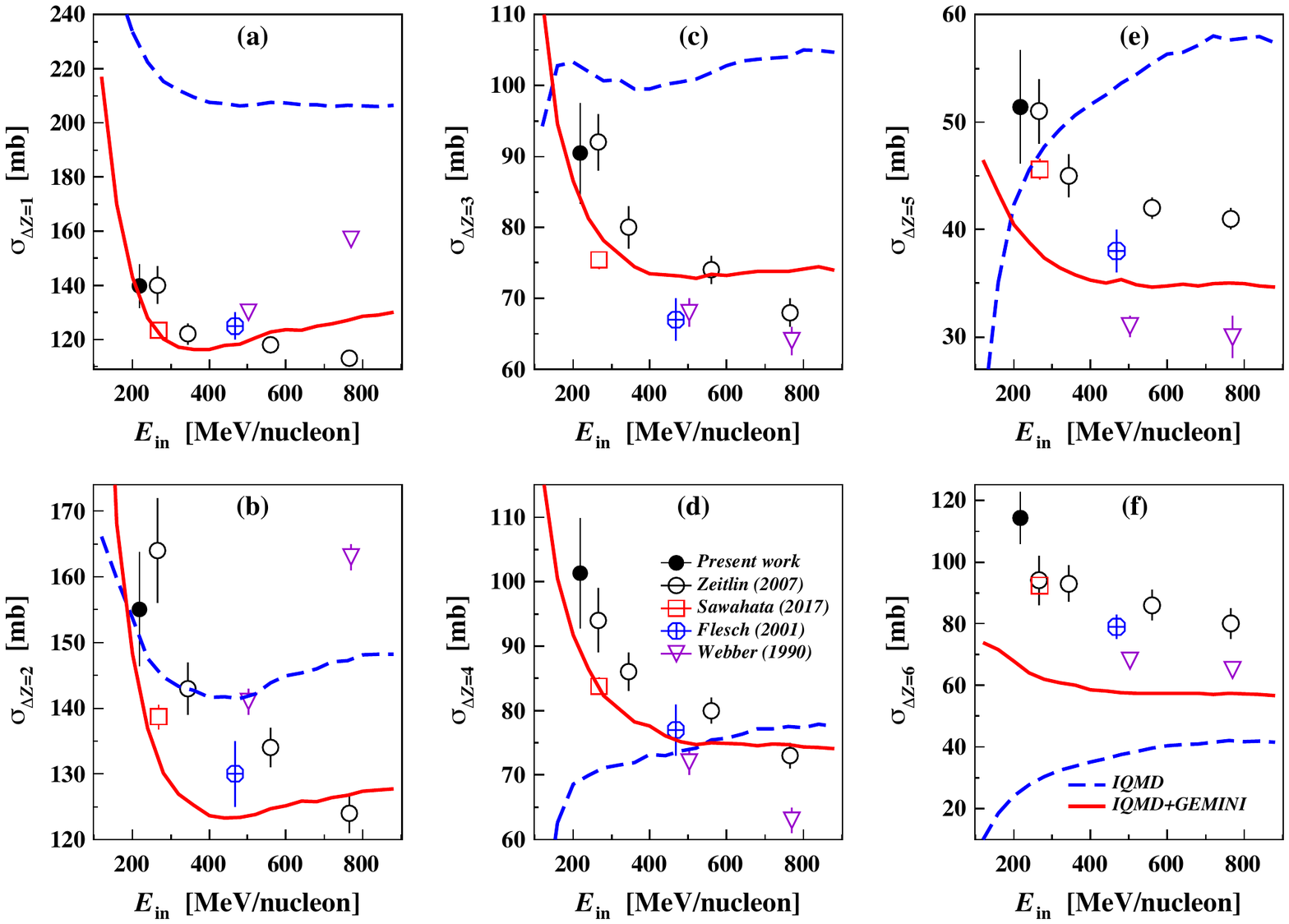}
    \caption{(Color online) EFCSs (symbols) as a function of incident energies for $^{28}$Si + $^{12}$C. The present data are shown by the full circles and the literature data represented by the open symbols are from Refs.~\cite{ZEITLIN2007341,SAWAHATA2017142,FLESCH2001237,PhysRevC.41.533}. 
    The IQMD and IQMD+GEMINI calculations are shown as dashed and solid lines, respectively.
    } 
    \label{fig5}
\end{figure*}

Figure~\ref{fig5} presents the EFCSs with $\Delta Z$ = 1-6 of $^{28}$Si fragments as a function of incident energies ranging from 200 to 800~MeV/nucleon~\cite{PhysRevC.41.533,FLESCH2001237,ZEITLIN2007341,SAWAHATA2017142}.
We estimate hereafter the total uncertainty of cross section as the square root of the sum of statistical and systematic uncertainties.
The data from CR-39 experiments~\cite{Li_2017,LI2016314} are not included here due to their relatively large uncertainties. 
In a global view, the data do not give 
a consistent energy dependence for $\Delta Z$ = 1 and 2.
In particular, 
the data reported by Webber $et$~$al.$~\cite{PhysRevC.41.533} and Zeitlin $et$~$al.$~\cite{ZEITLIN2007341} show an opposite trend 
with increasing energies.  
They also have a large difference in magnitude.
The data at 503 and 770~MeV/nucleon are systematically larger than those at 560 and 765~MeV/nucleon. 
As an example, the discrepancy at 770 MeV/nucleon can be up to about 40\% for $\Delta Z$ = 1.
On the other hand, the data for $\Delta Z$ = 3-6 tend to decrease with increasing incident energy, although the absolute values differ.
Both data obtained at 560 and 765~MeV/nucleon are systematically larger than those at 503 and 770 MeV/nucleon, which is the opposite of the case for $\Delta Z$ = 1-2.

The IQMD and IQMD+GEMINI calculations for 100 to 900 MeV/nucleon with a step of 40 MeV/nucleon are preformed in Fig.~\ref{fig5}, respectively.
The IQMD calculation only describes the first step, i.e., the formation of the excited prefragments.
Then, in the GEMINI model, highly excited residues will decay to the final fragments with lower $Z$ by emitting light particles (mainly including $p$, $n$, $d$, $\alpha$).
In general, the IQMD model predicts monotonically changing cross sections for $\Delta Z$ = 1-6.
The statistical deexcitation process in the GEMINI model accounts for 35-45\% reduction in cross sections for $\Delta Z$ = 1 and about 10\% for $\Delta Z$ = 2 above 200 MeV/nucleon, and changes the energy dependence for
$\Delta Z$ = 4-6. The IQMD+GEMINI results show a local minimum at about 300 MeV/nucleon for $\Delta Z$ = 1, which
is correlated to the energy dependence of nucleon-nucleon
cross section. For $\Delta Z$ = 2-6, the IQMD+GEMINI results decrease rapidly from 200 to 300 MeV/nucleon, and then appear almost constant above 300 MeV/nucleon.

\subsection{Comparisons with model predictions}

\begin{figure}[htpb!]
    \centering
    \includegraphics[width=0.48\textwidth]{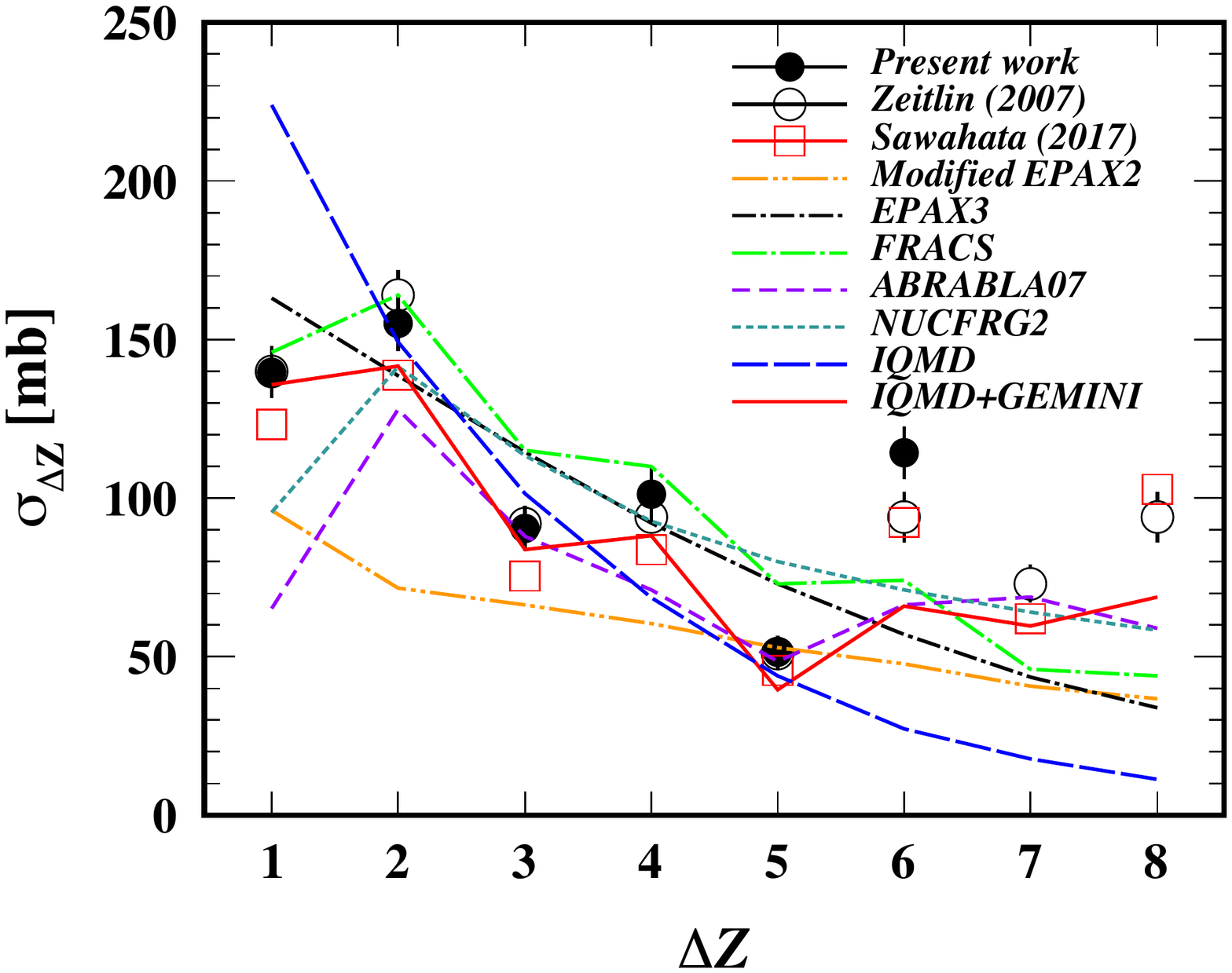}
    \caption{(Color online) EFCSs as a function of $\Delta Z$ for the fragments of $^{28}$Si on carbon. The solid circles are our experimental data. The open symbols are from literature~\cite{ZEITLIN2007341,SAWAHATA2017142}. For comparison, the model predictions are shown by different types of lines.}
    \label{fig6}
\end{figure}

Figure~\ref{fig6} presents the EFCSs of $^{28}$Si at 218, 266, and 268~MeV/nucleon along with model predictions. 
The EFCSs are getting overall smaller when the produced fragment is getting far away from the projectile.
One can see a clear odd-even staggering, i.e., an enhancement with even-$\textit{Z}$ isotope compared to that of the neighboring odd-$\textit{Z}$ isotope.
Both the modified EPAX2 and EPAX3 models predict a monotonically decreasing trend with increasing $\Delta Z$.
The FRACS results can generally follow the experimental trend for $\Delta Z$ = 1 to 6, 
but underestimate cross sections for $\Delta Z$ = 7 and 8.
For the abrasion-ablation models, the ABRABLA07 results can reproduce the experimental results fairly well.
The NUCFRG2 results show a monotonically decreasing trend with the increment of $\Delta Z$ from 2 to 8. 
Although both FRACS and ABRABLA07 models include the odd-even staggering, 
they predict decidedly different magnitudes of this staggering than what is exhibited by the data.

The IQMD+GEMINI calculation can reproduce the experimental data with an accuracy of better than 3.5\% except for $\Delta Z$ = 6 and 8.
It is worthwhile to note that the same model can also nicely reproduce cross sections measured in Refs.~\cite{IANCU2005525,PhysRevC.77.034605,PhysRevC.56.388}.
This demonstrates that the combination of primary dynamical collisions and sequential statistical deexcitation procedures can provide a reasonable description for heavy-ion collisions over a large range of incident energies.
For $\Delta Z$ = 6 and 8, a close comparison indicates that the IQMD+GEMINI calculation underestimates experimental data. This underestimation, however,
was not observed for data of 400~MeV/nucleon $^{36,40}$Al and 650~MeV/nucleon $^{35}$Cl on Al
target~\cite{IANCU2005525,PhysRevC.77.034605}.

The cross sections of primary fragments obtained by the IQMD model are compared to those of the final fragments by the IQMD+GEMINI model in Fig.~\ref{fig6}.
Differently from the IQMD+GEMINI results, the IQMD model predicts a monotonically decreasing cross sections with increasing $\Delta Z$ from 1 to 8;
this leads to a constant odd-even staggering strength. 
The high excitation energy of the primary fragments allows several times of sequential decay and hence huge decay paths.
The decay path ending as a final fragment with higher stability will be chosen with a higher probability. 

\subsection{Odd-even staggering}

\begin{figure}[htpb!]
    \centering
    \includegraphics[width=0.48\textwidth]{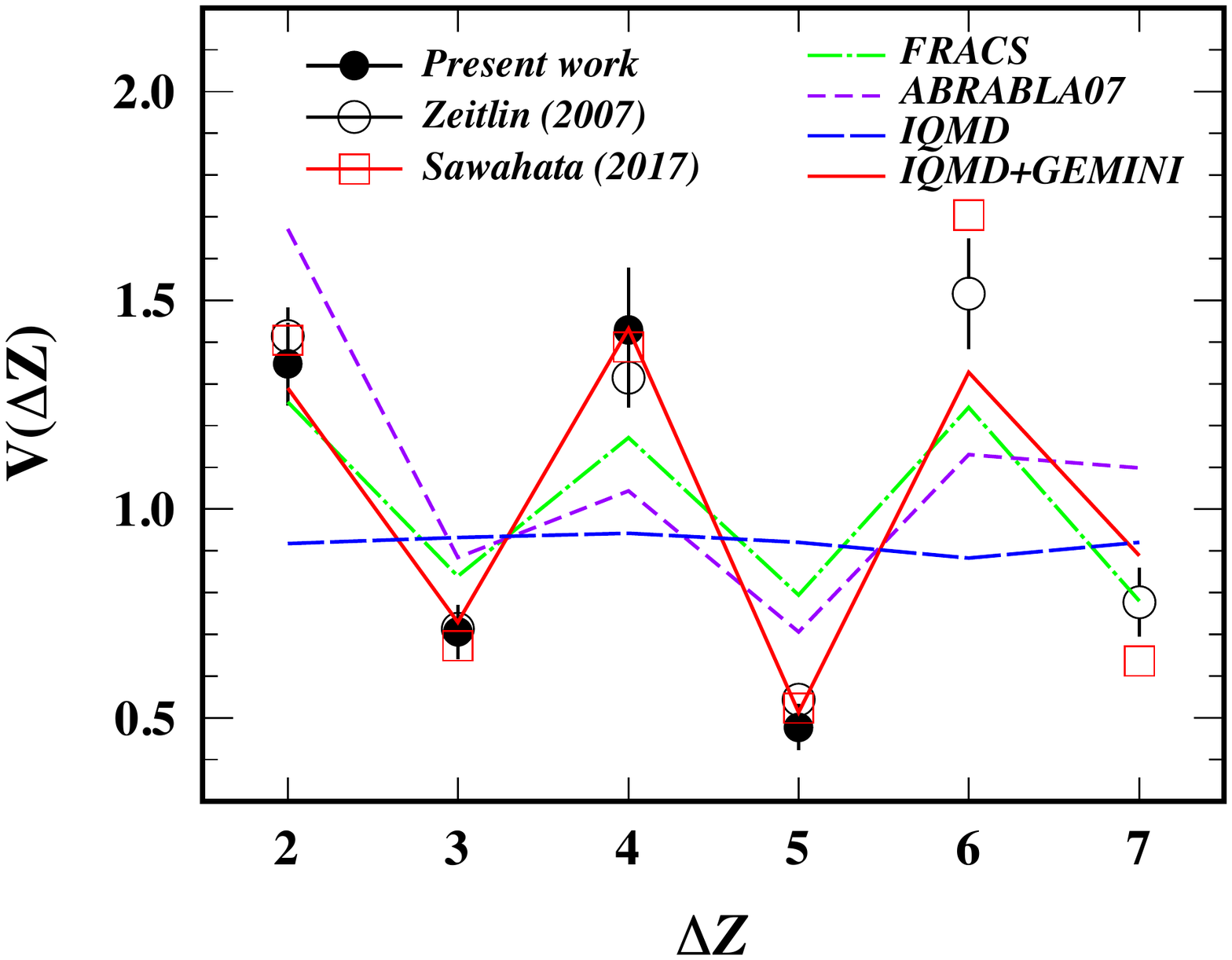}
    \caption{(Color online) Experimental $\textit{V}(\Delta Z)$ (symbols) as a function of $\Delta Z$ for fragment species of $^{28}$Si on carbon. The model predictions (lines) are shown for comparisons.}
    \label{fig7}
\end{figure}

The odd-even staggering has been discussed in the isotopic cross section from projectile fragmentation (see, e.g, in Refs.~\cite{RICCIARDI2004299,PhysRevC.97.044619,PhysRevC.105.064604}). 
To characterize the strength of odd-even staggering in charge distribution, the quantity $V(\Delta Z)$ is defined~\cite{IANCU2005525}:
\begin{equation}
    \textit{V}(\Delta Z)=\frac{2\sigma(\Delta Z)}{\sigma(\Delta Z+1)+\sigma(\Delta Z-1)},
\end{equation}
where the $\sigma(\Delta Z)$ refers to the EFCS with a particular knockout proton number $\Delta Z$.
Figure~\ref{fig7} presents $V(\Delta Z)$ as a function of $\Delta Z$ for $^{28}$Si fragments. Filled symbols are the data determined in the present work with $^{28}$Si at 218~MeV/nucleon, while open circles and open squares are those measured at 266~MeV/nucleon~\cite{ZEITLIN2007341} and 268~MeV/nucleon~\cite{SAWAHATA2017142}, respectively.
The $V(\Delta Z)$ values are larger than 1 for even-$Z$ and less than 1 for odd-$Z$ nuclei, and are close to 1 when odd-even staggering strength is weak.
In contrast, the IQMD+GEMINI results reproduce the experimental $V(\Delta Z)$ from $\Delta Z$ = 2 to 7, both in the strength and oscillation, while the ABRABLA07 and FRACS models predict a weaker staggering strength than experimental data.
It should be noted that the odd-even staggering strength $V(\Delta Z)$ will become weaker when using the cross sections determined with fixed widths in the fitting approach. The centroid values decrease by about 0.04 and 0.25 for $\Delta Z$ = 2 and 4, and increase by about 0.08 and 0.11 for $\Delta Z$ = 3 and 5, respectively.

Both the abrasion-ablation and IQMD+GEMINI are two-step models. Nevertheless, the calculated odd-even staggerings by those models are different, as shown in Fig.~\ref{fig7}.
To clarify the reasons, it is worthwhile to note that the treatments of the first step in the abrasion and IQMD models are distinctly different.
In the abrasion model, the number of abraded nucleons is proportional to the volume of the overlap region,
and the average excitation energy of the primary fragments after abrasion is estimated from the additional surface energy~\cite{PhysRevC.12.1888}.
This is a macroscopic treatment, but the treatment in the IQMD model is microscopic.
The calculations of the excitation energy of primary fragments depend on the dissipation of the incident energy and the energy-density functional, where not only the surface energy but also other interactions are considered.
The excitation energies of primary fragments by the IQMD model are larger than those by the abrasion model (see Fig.~2 in Ref.~\cite{PhysRevC.12.1888} and Fig.~5 in Ref.~\cite{PhysRevC.83.014608}).
The underestimation of the odd-even staggering by the abrasion-ablation models (ABRABLA07 and NUCFRG2) are
attributed to the low excitation energy of primary fragments.
Accordingly, in the decay of the primary fragments with low excitation energy, the diversity of the decay path is restrained and then the population of the final fragment depends more on the initial state rather than the stability of the final state. 

The odd-even staggering is attributed to the existence of pairing correlations in nuclear binding energy~\cite{RICCIARDI2004299}. 
In the IQMD+GEMINI model
which reproduces reasonably the odd-even staggering, nuclear masses and level density with pairing correlations are adopted in the GEMINI model.
On the other hand, pairing correlations are not taken into account in the IQMD model that treats the first step.
Experimental data in the reactions for the isotopic chains will be valuable to study the effect of the pairing correlation in the peripheral collisions.

\section{Summary}
\label{section5}

In summary, we reported new results on EFCSs of $^{28}$Si on carbon at 218 MeV/nucleon.
The data show an odd-even staggering with $\Delta Z$ = 1-6 and a local minimum at $\Delta Z$ = 5. 
Our experimental results are generally consistent with the previous measurements at similar energies.
We analyzed the dependence on the incident energy of fragment cross sections with $\Delta Z$ = 1-6.
By excluding the CR-39 results~\cite{LI2016314,Li_2017},
the existing data show in general a decreasing cross section for $\Delta Z$ = 3-6, but there is no decisive conclusion for $\Delta Z$ = 1 and 2.

In addition, the IQMD+GEMINI calculation of the EFCSs at different energies was performed. 
Deviations in both the magnitude and energy-dependent trend are seen when compared with the experimental data. This is partially due to the divergence in existing data.
High-quality data over different energy regions are thus called for to constrain this framework.

Among the various predictions performed, it is found that the modified EPAX2, EPAX3, and NUCFRG2 models cannot describe the EFCSs, while the ABRABLA07 and FRACS models predict a weaker odd-even staggering strength.
The IQMD+GEMINI model can reproduce the experimental data with an accuracy of better than 3.5\% for $\Delta Z\leq5$, both in the trend and the odd-even staggering. 
We then analyzed the formation mechanism of odd-even staggering in charge distribution by comparing the IQMD results with the IQMD+GEMINI.
The decay process from hot primary fragments in the GEMINI model shapes the final odd-even staggering.

\begin{acknowledgments}

We thank the HIRFL-CSR accelerator team for their efforts to provide a stable beam condition during the experiment. This work was supported partially by the National Natural Science Foundation of China (Grants No. U1832211, No. 11961141004, No. 11922501, No. 11475014, and No. 11905260) and the Western Light Project of the Chinese Academy of Sciences.

\section*{APPENDIX}
\appendix
\label{A}

The yield of one fragment (F) with an atomic number $Z$ within the target material can be expressed as the following differential equation~\cite{PhysRevC.66.014609}:
\setcounter{equation}{0}
\renewcommand\theequation{A\arabic{equation}}
\begin{equation}
\centering
    \frac{dN_{\rm{F}}(x)}{dx}=\sigma_{\Delta Z}^{\rm{P}}N_{0}(x)+\sum_{j>Z}N_{j}(x)\sigma^{j}_{\Delta Z}-\sigma^{\rm{F}}_{\rm{CC}}N_{\rm{F}}(x)\;,
    \label{eqA1}
\end{equation}
where the quantity $x$ denotes the target thickness traversed, expressed in nuclei per unit area. 
$\sigma_{\rm{CC}}^{\rm{F}}$ and $N_{\rm{F}}(x)$ represent the total charge-changing cross section of fragment F on target and the relevant count at the position $x$. 
$\Delta Z$ denotes the number of protons removed from the projectile.
$\sigma_{\Delta Z}^{\rm{P}}$ and $\sigma^{j}_{\Delta Z}$ are the relevant EFCSs of incident nuclide (P) and intermediate fragment $j$ with higher $Z$ with the target nuclides, while $N_{0}(x)$ and $N_{j}(x)$ are the counts of projectile and intermediate fragment $j$ at the position $x$, respectively.
The first term denotes the increase of fragment species with $Z$ produced from the incident beam bombarding the target, i.e., one-step reaction.
The second and the third terms represent the yield from intermediate fragments $j$ due to multistep reactions and the loss of the fragment species, respectively. 
Both are attributed to the secondary reactions in the target. Note that in Eq.~(\ref{eqA1}) the charge-exchange reaction is not considered since it is of typically less than mb.

The number of incident nuclei, $N_{0}(x)$, decays exponentially as a function of $x$ in the target:
\begin{equation}
    N_{0}(x)=N_{0}(0)\exp(-\sigma_{\rm{R}}^{\rm P}x)\;,
    \label{eqA2}
\end{equation}
where $N_{0}(0)$ and $\sigma^{\rm{P}}_{\rm{R}}$ are the total count and the reaction cross section of incident nuclei on target, respectively.
When the target thickness is comparable with the mean free path of the incident ions, the attenuation effect of incident beam and the secondary reactions in the target should be taken into account for a better accuracy.
In the following calculations, we describe our approach on 
how to deduce the EFCS and its uncertainty. This approach is general and can give a more precise
centroid value of the cross section.

\subsection{Thin target} 
When using a very thin target, the secondary reactions that contribute to the fragment of interest are negligible, $i.e.$, both the second and third terms on the right side of Eq.~(\ref{eqA1}) can be omitted. Together with Eq.~(\ref{eqA2}), $N_{\rm{F}}(x)$ is then determined as
\begin{equation}
  N_{\rm{F}}(x)=N_{0}(0)\frac{\sigma^{\rm{P}}_{\Delta Z}}{\sigma^{\rm{P}}_{\rm{R}}}[1-\exp(-\sigma^{\rm{P}}_{\rm{R}}x)]\;.
  \label{eqA3}
\end{equation}
In reality, to cancel the reactions induced by the materials other than the reaction target, practically, an alternative measurement on empty target can be carried out. Then the EFCS will be
\begin{equation}
    \sigma^{\rm{P}}_{\Delta Z}=\left(\frac{N_{\rm{F}}}{N_{0}}-\frac{N^{0}_{\rm{F}}}{N^{0}_{0}}\right)\frac{\sigma^{\rm{P}}_{\rm{R}}}{1-\exp(-\sigma^{\rm{P}}_{\rm{R}}t)}\;.
    \label{eqA4}
\end{equation}
In this experiment, the target thickness $t$ is given as 1.86 g/cm$^{2}$. The $\sigma^{\rm{P}}_{\rm{R}}$ is known with an accuracy of 5\%~\cite{PhysRevC.89.011601}, 
resulting in an uncertainty of 0.3\% in $\sigma^{\rm{P}}_{\Delta Z}$. 
It should be noted that Eq.~(\ref{eqA3}) can be simplified by expanding the exponential and neglecting the higher-order terms for an infinitely thin target, where the attenuation effect of incident beam can be safely neglected; then Eq.~(\ref{eqA3}) becomes

\begin{equation}
    N_{\rm{F}}(x)=N_{0}(0)\sigma^{\rm{P}}_{\Delta Z}x\;.
    \label{eqA5}
\end{equation}

\subsection{Thick target}
With the target thickness $x$ increasing, secondary reactions have to be considered as depicted in Eq.~(\ref{eqA1}). 
The incident nuclei may undergo cascade fragmentation process to produce the final fragment of interest, as shown by the second term of Eq.~(\ref{eqA1}). 
Here by omitting more than two-step cascade fragmentation process, the fragment yield from such two-step cascade fragmentation can be computed as follows:
\begin{equation}
\centering
    \frac{dN_{j}(x)}{dx}=N_{0}(x)\sigma^{\rm{P}}_{\Delta Z}-N_{j}(x)\sigma^{j}_{\rm{CC}}\;,
    \label{eqA6}
\end{equation}

\begin{equation}
\centering
    \frac{dN'_{\rm{F}}(x)}{dx}=N_{j}(x)\sigma^{j}_{\Delta Z}-N'_{\rm{F}}(x)\sigma^{\rm{F}}_{\rm{CC}}\;,
    \label{eqA7}
\end{equation}
where the $\sigma^{j}_{\rm{CC}}$ and $N_{j}(x)$ represent the total charge-changing cross section of intermediate fragment $j$ on target and the relevant count at the position $x$, respectively.
$N'_{\rm{F}}(x)$ is the number of final fragment produced by successive two-step fragmentation. 
The yield of the aimed fragment only from this two-step process is then
\begin{widetext}
\begin{equation}
\begin{aligned}
 N'_{\rm{F}}(x)=N_{0}(0)\sigma^{\rm{P}}_{\Delta Z}\sigma^{j}_{\Delta Z}\left[\frac{\exp(-\sigma^{\rm{P}}_{\rm{R}}x)}{(\sigma^{\rm{P}}_{\rm{R}}-\sigma^{j}_{\rm{CC}})(\sigma^{\rm{P}}_{\rm{R}}-\sigma^{\rm{F}}_{\rm{CC}})}+\frac{\exp(-\sigma^{\rm{F}}_{\rm{CC}}x)}{(\sigma^{\rm{F}}_{\rm{CC}}-\sigma^{\rm{P}}_{\rm{R}})(\sigma^{\rm{F}}_{\rm{CC}}-\sigma^{j}_{\rm{CC}})}+\frac{\exp(-\sigma^{j}_{\rm{CC}}x)}{(\sigma^{\rm{P}}_{\rm{R}}-\sigma^{j}_{\rm{CC}})(\sigma^{\rm{F}}_{\rm{CC}}-\sigma^{j}_{\rm{CC}})}
 \right].
    \label{eqA8}
\end{aligned}
\end{equation}
\end{widetext}
The total fragmentation yield is the sum of all possible paths to produce the final fragment, that is~\cite{BAZIN2002307}
\begin{equation}
    N_{\rm{F}}(x)= N^{1}_{\rm{F}}(x)+ \sum^{Z_{\rm{P}}}_{m=1}\sum^{N_{\rm{P}}}_{n=1}N'_{m,n,\rm{F}}(x)\;,
    \label{eqA9}
\end{equation}
where the $m$ and $n$ denote the proton and neutron numbers of the intermediate nuclei, respectively. $N^{1}_{\rm{F}}(x)$ is the relevant yield when considering the 
first and third terms on the right side of Eq.~(\ref{eqA1});
namely, the contributions from the second term are negligibly smaller compared to that from the first and third terms.

Let us then consider the special case in which the second term is neglected. $\Delta Z=1$ is one of such cases.
One can see that the second term accounts for about 8\% (0.4\%) of the third (first) term for $\Delta Z=1$ (see Table~\ref{tab2}).
Then Eq.~(\ref{eqA1}) can be rewritten as
\begin{equation}
\centering
    \frac{dN_{\rm{F}}(x)}{dx}=\sigma^{\rm{P}}_{\Delta Z}N_{0}(x)-\sigma^{\rm{F}}_{\rm{CC}}N_{\rm{F}}(x)\;.
    \label{eqA10}
\end{equation}
By inserting Eq.~(\ref{eqA2}) into Eq.~(\ref{eqA10}), one can solve the differential equation and obtain
\begin{equation}
\centering
N_{\rm{F}}(x)=\frac{N_{0}(0)\sigma^{\rm{P}}_{\Delta Z}\exp(-\sigma^{\rm{P}}_{\rm{R}}x)\{1-\exp[(\sigma^{\rm{P}}_{\rm{R}}-\sigma^{\rm{F}}_{\rm{CC}})x]\}}{\sigma^{\rm{F}}_{\rm{CC}}-\sigma^{\rm{P}}_{\rm{R}}}.
\label{eqA11}
\end{equation}
Experimentally, after subtracting the contributions from the materials other than the reaction target in the empty-target runs,
EFCS can be deduced as
\begin{equation}
\centering
\sigma^{\rm{P}}_{\Delta Z}=\left(\frac{N_{\rm{F}}}{N_{0}}-\frac{N^{0}_{\rm{F}}}{N^{0}_{0}}\right)\frac{\sigma^{\rm{F}}_{\rm{CC}}-\sigma^{\rm{P}}_{\rm{R}}}
{\exp(-\sigma^{\rm{P}}_{\rm{R}}t)-\exp(-\sigma^{\rm{F}}_{\rm{CC}}t)}\;. 
\label{eqA12}
\end{equation}
The total charge-changing cross section, $\sigma^{\rm{F}}_{\rm{CC}}$, can be calculated in the same framework as $\sigma^{\rm{P}}_{\rm{R}}$. 
The $\sigma^{\rm{F}}_{\rm{CC}}$ of 1120~mb for $\Delta Z=1$  
is determined by applying the correction of the scaling factor~\cite{PhysRevC.82.014609}.
The variances of 5\% of $\sigma^{\rm{P}}_{\rm{R}}$ and 15\% of $\sigma^{\rm{F}}_{\rm{CC}}$ will result in only about 0.9\% variance of $\sigma^{\rm{P}}_{\Delta Z}$.

To have a quantitative understanding, we take FRACS model as an example to calculate the cross section, $\sigma^{\rm{P}}_{\Delta Z}$, and obtain the relevant $N_{\rm{F}}$ from Eqs.~(\ref{eqA3}), (\ref{eqA5}), (\ref{eqA9}), and (\ref{eqA11}), respectively. 
Table~\ref{tab2} summarizes the calculated $N_{\rm{F}}$ of the elemental fragments with $\Delta Z$ = 1-6 for 218 MeV/nucleon $^{28}$Si on carbon target with a thickness of 1.86 g/$\rm{cm^{2}}$. 
One can see that
$N_{\rm{F}}$ deduced by Eq.~(\ref{eqA9}) represents the "true" EFCS, while $N_{\rm{F}}$ from (\ref{eqA11}) gives the lower limit to that from Eq.~(\ref{eqA9}). 
For smaller $\Delta Z$, $N_{\rm{F}}$ from Eq.~(\ref{eqA11}) is similar to the one from Eq.~(\ref{eqA9}) in magnitude.
This is due to the fact that the second term in Eq.~(\ref{eqA1}) (i.e., the yield from the higher-$Z$ fragment due to multistep reactions) plays a minor role. 
For larger $\Delta Z$, $N_{\rm{F}}$ from Eq.~(\ref{eqA3}) is almost identical to the one from Eq.~(\ref{eqA9}). 
This is due to the cancellation of contributions from the second and third terms in Eq.~(\ref{eqA1}). 
Moreover, the results from Eq.~(\ref{eqA5}) tend to be overestimated compared with those from Eqs.~(\ref{eqA3}) and (\ref{eqA9}).
Using Eq.~(\ref{eqA5}) to derive EFCS, as often done in EFCS determination, 
would therefore give an underestimated cross section, when the beam attenuation effect is significant.
We also examine the predictions by employing EPAX3, modified EPAX2, and Abrasion-Ablation models, and find that the above conclusions hold well, although the absolute values differ.

\textit{
As long as the absolute value of the second term is smaller than that of the third one in Eq.~(\ref{eqA1}),
the ``true" EFCS lies always in between those determined from Eqs.~(\ref{eqA3}) and (\ref{eqA11}), which set the lower and upper boundaries of EFCS, respectively.} 
The exception could occur at relatively large $\Delta Z$ and/or the thick target, as $\Delta Z$ = 6 in Table~\ref{tab2}.
To verify this, we compute $N_{\rm{F}}$ explicitly against the target thickness $x$ for $\Delta Z$ = 6, in which the contribution from the second term in Eq.~(\ref{eqA1}) could be more significant and be comparable to the third term.
Again, the FRACS model is employed to calculate the relevant cross sections. In Fig.~\ref{fig8}, one can see that $N_{\rm{F}}$ derived from Eqs.~(\ref{eqA3}), (\ref{eqA5}), (\ref{eqA9}), and (\ref{eqA11}) increases monotonically with increasing target thickness $x$.

\begin{figure}[htpb!]
    \centering
    \includegraphics[width=0.45\textwidth]{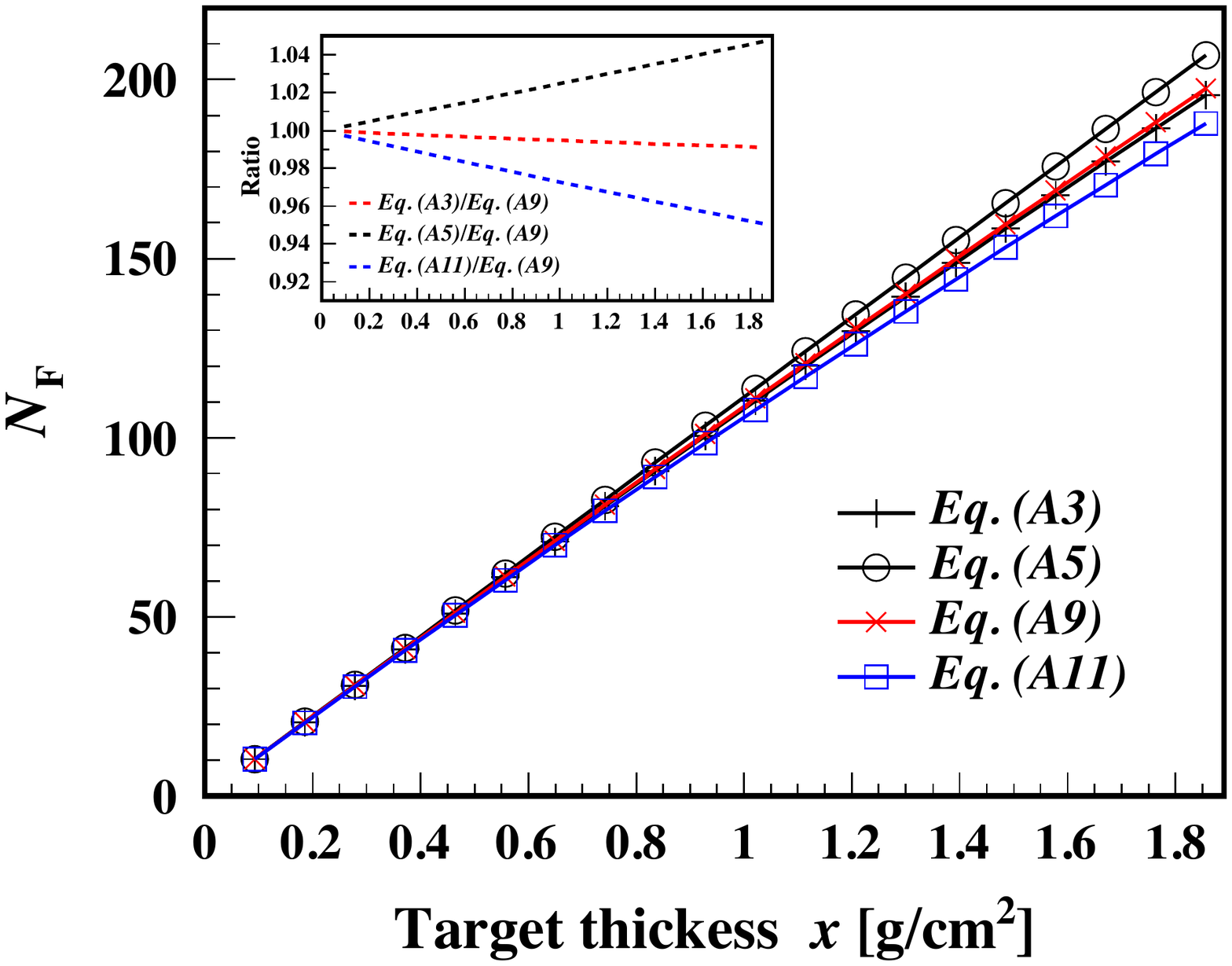}
    \caption{$N_{\rm{F}}$ as a function of target thickness $x$ for $\Delta Z$ = 6. Different symbols denote the $N_{\rm{F}}$ calculated by Eqs.~(\ref{eqA3}), (\ref{eqA5}), (\ref{eqA9}), and (\ref{eqA11}) within the framework of FRACS model. The ratios of $N_{\rm{F}}$ derived from Eqs.~(\ref{eqA3}), (\ref{eqA5}), and (\ref{eqA11}) to the one from Eq.~(\ref{eqA9}) are shown by red, black, and blue dotted lines in the inset, respectively.}
    \label{fig8}
\end{figure}

We take $N_{\rm{F}}$ derived from Eq.~(\ref{eqA9}) as a reference, and plot the ratios relative to $N_{\rm{F}}$ from Eqs.~(\ref{eqA3}),
(\ref{eqA5}), and (\ref{eqA11})  
in the inset of Fig.~\ref{fig8}.
Overall, the ratios of both Eq.~(\ref{eqA3})/Eq.~(\ref{eqA9}) and Eq.~(\ref{eqA11})/Eq.~(\ref{eqA9}) are less than 1 and decrease monotonically with the increment of target thickness $x$.
This indicates that $N_{\rm{F}}$ from Eq.~(\ref{eqA3}) is no longer a good upper limit for $\Delta Z=6$ when employing the FRACS predictions, although the deviation remains small.  
Quantitatively, for a very thin target, $N_{\rm{F}}$ deduced by Eqs.~(\ref{eqA3}), (\ref{eqA5}), and (\ref{eqA11}) have insignificant discrepancies from that by Eq.~(\ref{eqA9}).
When the target thickness is less than 0.3 g/cm$^2$, both the beam attenuation effect and secondary reactions in the target induce less than 1\% variance. 
In such case, Eq.~(\ref{eqA3}) or even Eq.~(\ref{eqA5}) can be used for the deduction of absolute cross sections with a good accuracy.
As the target thickness $x$ increases, however, the influence of these effects on $N_{\rm{F}}$ becomes significant.
When the target thickness is 1.86 g/$\rm{cm^{2}}$, the effect solely due to the beam attenuation effect in the target can result in a deviation of 5.3\%.
In addition to the beam attenuation effect,
neglecting either both the second and third terms or the second term in Eq.~(\ref{eqA1}) can lead to an additional deviation of 1\% or 5\% [see Eq.~(\ref{eqA3})/Eq.~(\ref{eqA9}) and Eq.~(\ref{eqA11})/Eq.~(\ref{eqA9})], respectively.

\subsection{EFCS and uncertainty} 

We found that Eq.~(\ref{eqA5}) is not a reliable way to deduce the EFCS, and a systematic overestimation  
up to 10\% can be introduced, in particular for a thick target and/or for large charge changes $\Delta Z$.
The final centroid values of EFCS are recommended to be calculated by Eq.~(\ref{eqA3}) for $\Delta Z$ = 1-2, 
by averaging the cross sections obtained from Eqs.~(\ref{eqA3}) and (\ref{eqA11}) for $\Delta Z$ = 3-4, and by Eq.~(\ref{eqA11}) for $\Delta Z$ = 5-6.
For the sake of reliability, the effects induced by the beam attenuation and secondary reactions in the target and by the data interpretation approaches are evaluated as systematic uncertainty. The former is taken as the half of the difference of cross sections from Eqs.~(\ref{eqA3}) and (\ref{eqA11}), while the latter is determined as half of the difference of cross sections in two different fitting methods for $\Delta Z$ = 1-6. Both are then added in quadrature to the systematic uncertainty, which is generally smaller than the statistical uncertainty. 
In reality, one should use Eqs.~(\ref{eqA4}) and (\ref{eqA12}) to remove the background events. 
Finally, the EFCS determination in the experimental condition is summarized in Table~\ref{tab3}.

\begin{table*}
    \centering
    \caption{$N_{\rm{F}}$ of the relevant elemental fragments with $\Delta Z$ from 1 to 6 for $^{28}$Si on the 1.86 g/cm$^2$ carbon. The FRACS model is used to calculate the cross section $\sigma^{\rm{P}}_{\Delta Z}$. The relevant $N_{\rm{F}}$ of fragment is obtained by using different formulas.
    }
   \setlength{\tabcolsep}{1.2mm}{
    \begin{tabular}{cccccccc}
    \hline
    \hline
     Model &  Formula & $N_{\textrm{F}(\Delta Z=1)}$ &  $N_{\textrm{F}(\Delta Z=2)}$ & $N_{\textrm{F}(\Delta Z=3)}$ & $N_{\textrm{F}(\Delta Z=4)}$ & $N_{\textrm{F}(\Delta Z=5)}$ & $N_{\textrm{F}(\Delta Z=6)}$\\
     \hline
    \multirow{4}{*}{FRACS}         &  Eq.~(\ref{eqA3})      & 491 & 449 & 491 & 315 & 249 & 196 \\
                                   &  Eq.~(\ref{eqA5})      & 518 & 474 & 518 & 333 & 269 & 207 \\
                                   &  Eq.~(\ref{eqA9})      & 468 & 431 & 474 & 309 & 247 & 198 \\ 
                                   &  Eq.~(\ref{eqA11})     & 466 & 427 & 467 & 300 & 237 & 188 \\
    \hline
    \hline
    \end{tabular}}
    \label{tab2}
\end{table*}

\begin{table}
    \centering
    \caption{Summary of relevant EFCS of fragments for $^{28}$Si + $^{12}$C at 218 MeV/nucleon for charge changes $\Delta Z$ from 1 to 6. The EFCSs derived from various formulas are listed in second, third, and fourth columns, respectively. The average of EFCSs from Eq.~(\ref{eqA4}) and Eq.~(\ref{eqA12}) is shown in the fifth column.
    The final centroid value of EFCS is given in the sixth column. The statistical and systematic uncertainties are presented in the seventh and eighth columns, respectively.}
    \setlength{\tabcolsep}{0.65mm}{
    \begin{tabular}{cccccccc}
    \hline
    \hline
    \multirow{2}{*}{$\Delta Z$} & $\sigma^{\rm{P}}_{\Delta Z}$\footnote{Equation~(\ref{eq1}).} & $\sigma^{\rm{P}}_{\Delta Z}$\footnote{Equation~(\ref{eqA4}).}  & $\sigma^{\rm{P}}_{\Delta Z}$\footnote{Equation~(\ref{eqA12}).} & $\sigma^{\rm{P(avg)}}_{\Delta Z}$ & $\sigma^{\rm{P(final)}}_{\Delta Z}$  & $\Delta\sigma^{\rm{P(stat)}}_{\Delta Z}$ & $\Delta\sigma^{\rm{P(sys)}}_{\Delta Z}$ \\
         & (mb) & (mb) & (mb) & (mb) & (mb) & (mb) & (mb) \\
    \hline     
     1  & 132.3  &  139.8  & 147.4  & 143.6 & 139.8  & 7.3  & 3.8 \\
     2  & 146.8  &  155.1  & 163.4  & 159.2 & 155.1  & 7.7  & 4.2 \\
     3  & 83.4   &  88.2   & 92.8   & 90.5  & 90.5   & 5.9  & 4.0 \\
     4  & 93.5   &  98.7   & 103.8  & 101.3 & 101.3  & 6.1  & 6.0 \\
     5  & 46.3   &  48.9   & 51.4   & 50.2  & 51.4   & 4.3  & 3.0 \\
     6  & 103.2  &  109.0  & 114.3  & 111.7 & 114.3  & 6.4  & 5.5 \\
    \hline
    \hline
    \end{tabular}}
    \label{tab3}
\end{table}

\end{acknowledgments}

\bibliographystyle{apsrev4-1}

\begin{thebibliography}{40}%
\makeatletter
\providecommand \@ifxundefined [1]{%
 \@ifx{#1\undefined}
}%
\providecommand \@ifnum [1]{%
 \ifnum #1\expandafter \@firstoftwo
 \else \expandafter \@secondoftwo
 \fi
}%
\providecommand \@ifx [1]{%
 \ifx #1\expandafter \@firstoftwo
 \else \expandafter \@secondoftwo
 \fi
}%
\providecommand \natexlab [1]{#1}%
\providecommand \enquote  [1]{``#1''}%
\providecommand \bibnamefont  [1]{#1}%
\providecommand \bibfnamefont [1]{#1}%
\providecommand \citenamefont [1]{#1}%
\providecommand \href@noop [0]{\@secondoftwo}%
\providecommand \href [0]{\begingroup \@sanitize@url \@href}%
\providecommand \@href[1]{\@@startlink{#1}\@@href}%
\providecommand \@@href[1]{\endgroup#1\@@endlink}%
\providecommand \@sanitize@url [0]{\catcode `\\12\catcode `\$12\catcode
  `\&12\catcode `\#12\catcode `\^12\catcode `\_12\catcode `\%12\relax}%
\providecommand \@@startlink[1]{}%
\providecommand \@@endlink[0]{}%
\providecommand \url  [0]{\begingroup\@sanitize@url \@url }%
\providecommand \@url [1]{\endgroup\@href {#1}{\urlprefix }}%
\providecommand \urlprefix  [0]{URL }%
\providecommand \Eprint [0]{\href }%
\providecommand \doibase [0]{http://dx.doi.org/}%
\providecommand \selectlanguage [0]{\@gobble}%
\providecommand \bibinfo  [0]{\@secondoftwo}%
\providecommand \bibfield  [0]{\@secondoftwo}%
\providecommand \translation [1]{[#1]}%
\providecommand \BibitemOpen [0]{}%
\providecommand \bibitemStop [0]{}%
\providecommand \bibitemNoStop [0]{.\EOS\space}%
\providecommand \EOS [0]{\spacefactor3000\relax}%
\providecommand \BibitemShut  [1]{\csname bibitem#1\endcsname}%
\let\auto@bib@innerbib\@empty
\bibitem [{\citenamefont {Ma}\ \emph {et~al.}(2021)\citenamefont {Ma},
  \citenamefont {Wei}, \citenamefont {Liu}, \citenamefont {Su}, \citenamefont
  {Zheng}, \citenamefont {Lin},\ and\ \citenamefont {Zhang}}]{MA2021103911}%
  \BibitemOpen
  \bibfield  {author} {\bibinfo {author} {\bibfnamefont {C.-W.}\ \bibnamefont
  {Ma}}, \bibinfo {author} {\bibfnamefont {H.-L.}\ \bibnamefont {Wei}},
  \bibinfo {author} {\bibfnamefont {X.-Q.}\ \bibnamefont {Liu}}, \bibinfo
  {author} {\bibfnamefont {J.}~\bibnamefont {Su}}, \bibinfo {author}
  {\bibfnamefont {H.}~\bibnamefont {Zheng}}, \bibinfo {author} {\bibfnamefont
  {W.-P.}\ \bibnamefont {Lin}}, \ and\ \bibinfo {author} {\bibfnamefont
  {Y.-X.}\ \bibnamefont {Zhang}},\ }\href {\doibase
  https://doi.org/10.1016/j.ppnp.2021.103911} {\bibfield  {journal} {\bibinfo
  {journal} {Prog. Part. Nucl. Phys.}\ }\textbf {\bibinfo {volume} {121}},\
  \bibinfo {pages} {103911} (\bibinfo {year} {2021})}\BibitemShut {NoStop}%
\bibitem [{\citenamefont {Durante}\ and\ \citenamefont
  {Cucinotta}(2011)}]{RevModPhys.83.1245}%
  \BibitemOpen
  \bibfield  {author} {\bibinfo {author} {\bibfnamefont {M.}~\bibnamefont
  {Durante}}\ and\ \bibinfo {author} {\bibfnamefont {F.~A.}\ \bibnamefont
  {Cucinotta}},\ }\href {\doibase 10.1103/RevModPhys.83.1245} {\bibfield
  {journal} {\bibinfo  {journal} {Rev. Mod. Phys.}\ }\textbf {\bibinfo {volume}
  {83}},\ \bibinfo {pages} {1245} (\bibinfo {year} {2011})}\BibitemShut
  {NoStop}%
\bibitem [{\citenamefont {Webber}\ \emph
  {et~al.}(1990{\natexlab{a}})\citenamefont {Webber}, \citenamefont {Kish},\
  and\ \citenamefont {Schrier}}]{PhysRevC.41.533}%
  \BibitemOpen
  \bibfield  {author} {\bibinfo {author} {\bibfnamefont {W.~R.}\ \bibnamefont
  {Webber}}, \bibinfo {author} {\bibfnamefont {J.~C.}\ \bibnamefont {Kish}}, \
  and\ \bibinfo {author} {\bibfnamefont {D.~A.}\ \bibnamefont {Schrier}},\
  }\href {\doibase 10.1103/PhysRevC.41.533} {\bibfield  {journal} {\bibinfo
  {journal} {Phys. Rev. C}\ }\textbf {\bibinfo {volume} {41}},\ \bibinfo
  {pages} {533} (\bibinfo {year} {1990}{\natexlab{a}})}\BibitemShut {NoStop}%
\bibitem [{\citenamefont {Flesch}\ \emph {et~al.}(2001)\citenamefont {Flesch},
  \citenamefont {Iancu}, \citenamefont {Heinrich},\ and\ \citenamefont
  {Yasuda}}]{FLESCH2001237}%
  \BibitemOpen
  \bibfield  {author} {\bibinfo {author} {\bibfnamefont {F.}~\bibnamefont
  {Flesch}}, \bibinfo {author} {\bibfnamefont {G.}~\bibnamefont {Iancu}},
  \bibinfo {author} {\bibfnamefont {W.}~\bibnamefont {Heinrich}}, \ and\
  \bibinfo {author} {\bibfnamefont {H.}~\bibnamefont {Yasuda}},\ }\href
  {\doibase https://doi.org/10.1016/S1350-4487(01)00158-5} {\bibfield
  {journal} {\bibinfo  {journal} {Radiat. Meas.}\ }\textbf {\bibinfo {volume}
  {34}},\ \bibinfo {pages} {237} (\bibinfo {year} {2001})}\BibitemShut
  {NoStop}%
\bibitem [{\citenamefont {Zeitlin}\ \emph {et~al.}(2007)\citenamefont
  {Zeitlin}, \citenamefont {Fukumura}, \citenamefont {Guetersloh},
  \citenamefont {Heilbronn}, \citenamefont {Iwata}, \citenamefont {Miller},\
  and\ \citenamefont {Murakami}}]{ZEITLIN2007341}%
  \BibitemOpen
  \bibfield  {author} {\bibinfo {author} {\bibfnamefont {C.}~\bibnamefont
  {Zeitlin}}, \bibinfo {author} {\bibfnamefont {A.}~\bibnamefont {Fukumura}},
  \bibinfo {author} {\bibfnamefont {S.}~\bibnamefont {Guetersloh}}, \bibinfo
  {author} {\bibfnamefont {L.}~\bibnamefont {Heilbronn}}, \bibinfo {author}
  {\bibfnamefont {Y.}~\bibnamefont {Iwata}}, \bibinfo {author} {\bibfnamefont
  {J.}~\bibnamefont {Miller}}, \ and\ \bibinfo {author} {\bibfnamefont
  {T.}~\bibnamefont {Murakami}},\ }\href {\doibase
  https://doi.org/10.1016/j.nuclphysa.2006.10.088} {\bibfield  {journal}
  {\bibinfo  {journal} {Nucl. Phys. A}\ }\textbf {\bibinfo {volume} {784}},\
  \bibinfo {pages} {341} (\bibinfo {year} {2007})}\BibitemShut {NoStop}%
\bibitem [{\citenamefont {Cecchini}\ \emph {et~al.}(2008)\citenamefont
  {Cecchini}, \citenamefont {Chiarusi}, \citenamefont {Giacomelli},
  \citenamefont {Giorgini}, \citenamefont {Kumar}, \citenamefont {Mandrioli},
  \citenamefont {Manzoor}, \citenamefont {Margiotta}, \citenamefont
  {Medinaceli}, \citenamefont {Patrizii}, \citenamefont {Popa}, \citenamefont
  {Qureshi}, \citenamefont {Sirri}, \citenamefont {Spurio},\ and\ \citenamefont
  {Togo}}]{CECCHINI2008206}%
  \BibitemOpen
  \bibfield  {author} {\bibinfo {author} {\bibfnamefont {S.}~\bibnamefont
  {Cecchini}}, \bibinfo {author} {\bibfnamefont {T.}~\bibnamefont {Chiarusi}},
  \bibinfo {author} {\bibfnamefont {G.}~\bibnamefont {Giacomelli}}, \bibinfo
  {author} {\bibfnamefont {M.}~\bibnamefont {Giorgini}}, \bibinfo {author}
  {\bibfnamefont {A.}~\bibnamefont {Kumar}}, \bibinfo {author} {\bibfnamefont
  {G.}~\bibnamefont {Mandrioli}}, \bibinfo {author} {\bibfnamefont
  {S.}~\bibnamefont {Manzoor}}, \bibinfo {author} {\bibfnamefont
  {A.}~\bibnamefont {Margiotta}}, \bibinfo {author} {\bibfnamefont
  {E.}~\bibnamefont {Medinaceli}}, \bibinfo {author} {\bibfnamefont
  {L.}~\bibnamefont {Patrizii}}, \bibinfo {author} {\bibfnamefont
  {V.}~\bibnamefont {Popa}}, \bibinfo {author} {\bibfnamefont {I.}~\bibnamefont
  {Qureshi}}, \bibinfo {author} {\bibfnamefont {G.}~\bibnamefont {Sirri}},
  \bibinfo {author} {\bibfnamefont {M.}~\bibnamefont {Spurio}}, \ and\ \bibinfo
  {author} {\bibfnamefont {V.}~\bibnamefont {Togo}},\ }\href {\doibase
  https://doi.org/10.1016/j.nuclphysa.2008.03.017} {\bibfield  {journal}
  {\bibinfo  {journal} {Nucl. Phys. A}\ }\textbf {\bibinfo {volume} {807}},\
  \bibinfo {pages} {206} (\bibinfo {year} {2008})}\BibitemShut {NoStop}%
\bibitem [{\citenamefont {Sawahata}\ \emph {et~al.}(2017)\citenamefont
  {Sawahata}, \citenamefont {Ozawa}, \citenamefont {Saito}, \citenamefont
  {Abe}, \citenamefont {Ichikawa}, \citenamefont {Inaba}, \citenamefont
  {Ishibashi}, \citenamefont {Kitagawa}, \citenamefont {Matsunaga},
  \citenamefont {Moriguchi}, \citenamefont {Nagae}, \citenamefont {Okada},
  \citenamefont {Sato}, \citenamefont {Suzuki}, \citenamefont {Suzuki},
  \citenamefont {Takeuchi}, \citenamefont {Yamaguchi},\ and\ \citenamefont
  {Zenihiro}}]{SAWAHATA2017142}%
  \BibitemOpen
  \bibfield  {author} {\bibinfo {author} {\bibfnamefont {K.}~\bibnamefont
  {Sawahata}}, \bibinfo {author} {\bibfnamefont {A.}~\bibnamefont {Ozawa}},
  \bibinfo {author} {\bibfnamefont {Y.}~\bibnamefont {Saito}}, \bibinfo
  {author} {\bibfnamefont {Y.}~\bibnamefont {Abe}}, \bibinfo {author}
  {\bibfnamefont {Y.}~\bibnamefont {Ichikawa}}, \bibinfo {author}
  {\bibfnamefont {N.}~\bibnamefont {Inaba}}, \bibinfo {author} {\bibfnamefont
  {Y.}~\bibnamefont {Ishibashi}}, \bibinfo {author} {\bibfnamefont
  {A.}~\bibnamefont {Kitagawa}}, \bibinfo {author} {\bibfnamefont
  {S.}~\bibnamefont {Matsunaga}}, \bibinfo {author} {\bibfnamefont
  {T.}~\bibnamefont {Moriguchi}}, \bibinfo {author} {\bibfnamefont
  {D.}~\bibnamefont {Nagae}}, \bibinfo {author} {\bibfnamefont
  {S.}~\bibnamefont {Okada}}, \bibinfo {author} {\bibfnamefont
  {S.}~\bibnamefont {Sato}}, \bibinfo {author} {\bibfnamefont {S.}~\bibnamefont
  {Suzuki}}, \bibinfo {author} {\bibfnamefont {T.}~\bibnamefont {Suzuki}},
  \bibinfo {author} {\bibfnamefont {Y.}~\bibnamefont {Takeuchi}}, \bibinfo
  {author} {\bibfnamefont {T.}~\bibnamefont {Yamaguchi}}, \ and\ \bibinfo
  {author} {\bibfnamefont {J.}~\bibnamefont {Zenihiro}},\ }\href {\doibase
  https://doi.org/10.1016/j.nuclphysa.2017.02.012} {\bibfield  {journal}
  {\bibinfo  {journal} {Nucl. Phys. A}\ }\textbf {\bibinfo {volume} {961}},\
  \bibinfo {pages} {142} (\bibinfo {year} {2017})}\BibitemShut {NoStop}%
\bibitem [{\citenamefont {Li}\ \emph {et~al.}(2016)\citenamefont {Li},
  \citenamefont {Zhang}, \citenamefont {Cheng}, \citenamefont {Kodaira},\ and\
  \citenamefont {Yasuda}}]{LI2016314}%
  \BibitemOpen
  \bibfield  {author} {\bibinfo {author} {\bibfnamefont {J.-S.}\ \bibnamefont
  {Li}}, \bibinfo {author} {\bibfnamefont {D.-H.}\ \bibnamefont {Zhang}},
  \bibinfo {author} {\bibfnamefont {J.-X.}\ \bibnamefont {Cheng}}, \bibinfo
  {author} {\bibfnamefont {S.}~\bibnamefont {Kodaira}}, \ and\ \bibinfo
  {author} {\bibfnamefont {N.}~\bibnamefont {Yasuda}},\ }\href {\doibase
  https://doi.org/10.1016/j.cjph.2016.05.003} {\bibfield  {journal} {\bibinfo
  {journal} {Chin. J. Phys.}\ }\textbf {\bibinfo {volume} {54}},\ \bibinfo
  {pages} {314} (\bibinfo {year} {2016})}\BibitemShut {NoStop}%
\bibitem [{\citenamefont {Li}\ \emph {et~al.}(2017)\citenamefont {Li},
  \citenamefont {Dang}, \citenamefont {Zhang}, \citenamefont {Cheng},
  \citenamefont {Kodaira},\ and\ \citenamefont {Yasuda}}]{Li_2017}%
  \BibitemOpen
  \bibfield  {author} {\bibinfo {author} {\bibfnamefont {J.-S.}\ \bibnamefont
  {Li}}, \bibinfo {author} {\bibfnamefont {Y.-H.}\ \bibnamefont {Dang}},
  \bibinfo {author} {\bibfnamefont {D.-H.}\ \bibnamefont {Zhang}}, \bibinfo
  {author} {\bibfnamefont {J.-X.}\ \bibnamefont {Cheng}}, \bibinfo {author}
  {\bibfnamefont {S.}~\bibnamefont {Kodaira}}, \ and\ \bibinfo {author}
  {\bibfnamefont {N.}~\bibnamefont {Yasuda}},\ }\href {\doibase
  10.1088/0256-307x/34/10/102501} {\bibfield  {journal} {\bibinfo  {journal}
  {Chin. Phys. Lett.}\ }\textbf {\bibinfo {volume} {34}},\ \bibinfo {pages}
  {102501} (\bibinfo {year} {2017})}\BibitemShut {NoStop}%
\bibitem [{\citenamefont {Su}\ \emph {et~al.}(2011)\citenamefont {Su},
  \citenamefont {Zhang},\ and\ \citenamefont {Bian}}]{PhysRevC.83.014608}%
  \BibitemOpen
  \bibfield  {author} {\bibinfo {author} {\bibfnamefont {J.}~\bibnamefont
  {Su}}, \bibinfo {author} {\bibfnamefont {F.-S.}\ \bibnamefont {Zhang}}, \
  and\ \bibinfo {author} {\bibfnamefont {B.-A.}\ \bibnamefont {Bian}},\ }\href
  {\doibase 10.1103/PhysRevC.83.014608} {\bibfield  {journal} {\bibinfo
  {journal} {Phys. Rev. C}\ }\textbf {\bibinfo {volume} {83}},\ \bibinfo
  {pages} {014608} (\bibinfo {year} {2011})}\BibitemShut {NoStop}%
\bibitem [{\citenamefont {Cheng}\ \emph {et~al.}(2012)\citenamefont {Cheng},
  \citenamefont {Jiang}, \citenamefont {Yan},\ and\ \citenamefont
  {Zhang}}]{Cheng_2012}%
  \BibitemOpen
  \bibfield  {author} {\bibinfo {author} {\bibfnamefont {J.~X.}\ \bibnamefont
  {Cheng}}, \bibinfo {author} {\bibfnamefont {X.}~\bibnamefont {Jiang}},
  \bibinfo {author} {\bibfnamefont {S.~W.}\ \bibnamefont {Yan}}, \ and\
  \bibinfo {author} {\bibfnamefont {D.~H.}\ \bibnamefont {Zhang}},\ }\href
  {\doibase 10.1088/0954-3899/39/5/055104} {\bibfield  {journal} {\bibinfo
  {journal} {J. Phys. G: Nucl. Part. Phys.}\ }\textbf {\bibinfo {volume}
  {39}},\ \bibinfo {pages} {055104} (\bibinfo {year} {2012})}\BibitemShut
  {NoStop}%
\bibitem [{\citenamefont {Cheng}\ \emph {et~al.}(2015)\citenamefont {Cheng},
  \citenamefont {Tian},\ and\ \citenamefont {Zhang}}]{Cheng_2015}%
  \BibitemOpen
  \bibfield  {author} {\bibinfo {author} {\bibfnamefont {J.~X.}\ \bibnamefont
  {Cheng}}, \bibinfo {author} {\bibfnamefont {J.~L.}\ \bibnamefont {Tian}}, \
  and\ \bibinfo {author} {\bibfnamefont {D.~H.}\ \bibnamefont {Zhang}},\ }\href
  {\doibase 10.1088/0954-3899/42/1/015102} {\bibfield  {journal} {\bibinfo
  {journal} {J. Phys. G: Nucl. Part. Phys.}\ }\textbf {\bibinfo {volume}
  {42}},\ \bibinfo {pages} {015102} (\bibinfo {year} {2015})}\BibitemShut
  {NoStop}%
\bibitem [{\citenamefont {S\"ummerer}(2012)}]{PhysRevC.86.014601}%
  \BibitemOpen
  \bibfield  {author} {\bibinfo {author} {\bibfnamefont {K.}~\bibnamefont
  {S\"ummerer}},\ }\href {\doibase 10.1103/PhysRevC.86.014601} {\bibfield
  {journal} {\bibinfo  {journal} {Phys. Rev. C}\ }\textbf {\bibinfo {volume}
  {86}},\ \bibinfo {pages} {014601} (\bibinfo {year} {2012})}\BibitemShut
  {NoStop}%
\bibitem [{\citenamefont {Zhang}(2013)}]{ZHANG201359}%
  \BibitemOpen
  \bibfield  {author} {\bibinfo {author} {\bibfnamefont {X.}~\bibnamefont
  {Zhang}},\ }\href {\doibase https://doi.org/10.1016/j.nuclphysa.2013.06.013}
  {\bibfield  {journal} {\bibinfo  {journal} {Nucl. Phys. A}\ }\textbf
  {\bibinfo {volume} {915}},\ \bibinfo {pages} {59} (\bibinfo {year}
  {2013})}\BibitemShut {NoStop}%
\bibitem [{\citenamefont {Mei}(2017)}]{PhysRevC.95.034608}%
  \BibitemOpen
  \bibfield  {author} {\bibinfo {author} {\bibfnamefont {B.}~\bibnamefont
  {Mei}},\ }\href {\doibase 10.1103/PhysRevC.95.034608} {\bibfield  {journal}
  {\bibinfo  {journal} {Phys. Rev. C}\ }\textbf {\bibinfo {volume} {95}},\
  \bibinfo {pages} {034608} (\bibinfo {year} {2017})}\BibitemShut {NoStop}%
\bibitem [{\citenamefont {Wilson}\ \emph {et~al.}(1994)\citenamefont {Wilson},
  \citenamefont {Shinn}, \citenamefont {Townsend}, \citenamefont {Tripathi},
  \citenamefont {Badavi},\ and\ \citenamefont {Chun}}]{WILSON199495}%
  \BibitemOpen
  \bibfield  {author} {\bibinfo {author} {\bibfnamefont {J.}~\bibnamefont
  {Wilson}}, \bibinfo {author} {\bibfnamefont {J.}~\bibnamefont {Shinn}},
  \bibinfo {author} {\bibfnamefont {L.}~\bibnamefont {Townsend}}, \bibinfo
  {author} {\bibfnamefont {R.}~\bibnamefont {Tripathi}}, \bibinfo {author}
  {\bibfnamefont {F.}~\bibnamefont {Badavi}}, \ and\ \bibinfo {author}
  {\bibfnamefont {S.}~\bibnamefont {Chun}},\ }\href {\doibase
  https://doi.org/10.1016/0168-583X(94)95662-6} {\bibfield  {journal} {\bibinfo
   {journal} {Nucl. Instrum. Methods Phys. Res., Sect. B}\ }\textbf {\bibinfo
  {volume} {94}},\ \bibinfo {pages} {95} (\bibinfo {year} {1994})}\BibitemShut
  {NoStop}%
\bibitem [{\citenamefont {H\"ufner}\ \emph {et~al.}(1975)\citenamefont
  {H\"ufner}, \citenamefont {Sch\"afer},\ and\ \citenamefont
  {Sch\"urmann}}]{PhysRevC.12.1888}%
  \BibitemOpen
  \bibfield  {author} {\bibinfo {author} {\bibfnamefont {J.}~\bibnamefont
  {H\"ufner}}, \bibinfo {author} {\bibfnamefont {K.}~\bibnamefont {Sch\"afer}},
  \ and\ \bibinfo {author} {\bibfnamefont {B.}~\bibnamefont {Sch\"urmann}},\
  }\href {\doibase 10.1103/PhysRevC.12.1888} {\bibfield  {journal} {\bibinfo
  {journal} {Phys. Rev. C}\ }\textbf {\bibinfo {volume} {12}},\ \bibinfo
  {pages} {1888} (\bibinfo {year} {1975})}\BibitemShut {NoStop}%
\bibitem [{\citenamefont {Hartnack}\ \emph {et~al.}(1989)\citenamefont
  {Hartnack}, \citenamefont {Zhuxia}, \citenamefont {Neise}, \citenamefont
  {Peilert}, \citenamefont {Rosenhauer}, \citenamefont {Sorge}, \citenamefont
  {Aichelin}, \citenamefont {Stöcker},\ and\ \citenamefont
  {Greiner}}]{HARTNACK1989303}%
  \BibitemOpen
  \bibfield  {author} {\bibinfo {author} {\bibfnamefont {C.}~\bibnamefont
  {Hartnack}}, \bibinfo {author} {\bibfnamefont {L.}~\bibnamefont {Zhuxia}},
  \bibinfo {author} {\bibfnamefont {L.}~\bibnamefont {Neise}}, \bibinfo
  {author} {\bibfnamefont {G.}~\bibnamefont {Peilert}}, \bibinfo {author}
  {\bibfnamefont {A.}~\bibnamefont {Rosenhauer}}, \bibinfo {author}
  {\bibfnamefont {H.}~\bibnamefont {Sorge}}, \bibinfo {author} {\bibfnamefont
  {J.}~\bibnamefont {Aichelin}}, \bibinfo {author} {\bibfnamefont
  {H.}~\bibnamefont {Stöcker}}, \ and\ \bibinfo {author} {\bibfnamefont
  {W.}~\bibnamefont {Greiner}},\ }\href {\doibase
  https://doi.org/10.1016/0375-9474(89)90328-X} {\bibfield  {journal} {\bibinfo
   {journal} {Nucl. Phys. A}\ }\textbf {\bibinfo {volume} {495}},\ \bibinfo
  {pages} {303} (\bibinfo {year} {1989})}\BibitemShut {NoStop}%
\bibitem [{\citenamefont {Charity}\ \emph {et~al.}(1988)\citenamefont
  {Charity}, \citenamefont {McMahan}, \citenamefont {Wozniak}, \citenamefont
  {McDonald}, \citenamefont {Moretto}, \citenamefont {Sarantites},
  \citenamefont {Sobotka}, \citenamefont {Guarino}, \citenamefont {Pantaleo},
  \citenamefont {Fiore}, \citenamefont {Gobbi},\ and\ \citenamefont
  {Hildenbrand}}]{CHARITY1988371}%
  \BibitemOpen
  \bibfield  {author} {\bibinfo {author} {\bibfnamefont {R.}~\bibnamefont
  {Charity}}, \bibinfo {author} {\bibfnamefont {M.}~\bibnamefont {McMahan}},
  \bibinfo {author} {\bibfnamefont {G.}~\bibnamefont {Wozniak}}, \bibinfo
  {author} {\bibfnamefont {R.}~\bibnamefont {McDonald}}, \bibinfo {author}
  {\bibfnamefont {L.}~\bibnamefont {Moretto}}, \bibinfo {author} {\bibfnamefont
  {D.}~\bibnamefont {Sarantites}}, \bibinfo {author} {\bibfnamefont
  {L.}~\bibnamefont {Sobotka}}, \bibinfo {author} {\bibfnamefont
  {G.}~\bibnamefont {Guarino}}, \bibinfo {author} {\bibfnamefont
  {A.}~\bibnamefont {Pantaleo}}, \bibinfo {author} {\bibfnamefont
  {L.}~\bibnamefont {Fiore}}, \bibinfo {author} {\bibfnamefont
  {A.}~\bibnamefont {Gobbi}}, \ and\ \bibinfo {author} {\bibfnamefont
  {K.}~\bibnamefont {Hildenbrand}},\ }\href {\doibase
  https://doi.org/10.1016/0375-9474(88)90542-8} {\bibfield  {journal} {\bibinfo
   {journal} {Nucl. Phys. A}\ }\textbf {\bibinfo {volume} {483}},\ \bibinfo
  {pages} {371} (\bibinfo {year} {1988})}\BibitemShut {NoStop}%
\bibitem [{\citenamefont {Xia}\ \emph {et~al.}(2002)\citenamefont {Xia},
  \citenamefont {Zhan}, \citenamefont {Wei}, \citenamefont {Yuan},
  \citenamefont {Song}, \citenamefont {Zhang}, \citenamefont {Yang},
  \citenamefont {Yuan}, \citenamefont {Gao}, \citenamefont {Zhao},
  \citenamefont {Yang}, \citenamefont {Xiao}, \citenamefont {Man},
  \citenamefont {Dang}, \citenamefont {Cai}, \citenamefont {Wang},
  \citenamefont {Tang}, \citenamefont {Qiao}, \citenamefont {Rao},
  \citenamefont {He}, \citenamefont {Mao},\ and\ \citenamefont
  {Zhou}}]{XIA200211}%
  \BibitemOpen
  \bibfield  {author} {\bibinfo {author} {\bibfnamefont {J.}~\bibnamefont
  {Xia}}, \bibinfo {author} {\bibfnamefont {W.}~\bibnamefont {Zhan}}, \bibinfo
  {author} {\bibfnamefont {B.}~\bibnamefont {Wei}}, \bibinfo {author}
  {\bibfnamefont {Y.}~\bibnamefont {Yuan}}, \bibinfo {author} {\bibfnamefont
  {M.}~\bibnamefont {Song}}, \bibinfo {author} {\bibfnamefont {W.}~\bibnamefont
  {Zhang}}, \bibinfo {author} {\bibfnamefont {X.}~\bibnamefont {Yang}},
  \bibinfo {author} {\bibfnamefont {P.}~\bibnamefont {Yuan}}, \bibinfo {author}
  {\bibfnamefont {D.}~\bibnamefont {Gao}}, \bibinfo {author} {\bibfnamefont
  {H.}~\bibnamefont {Zhao}}, \bibinfo {author} {\bibfnamefont {X.}~\bibnamefont
  {Yang}}, \bibinfo {author} {\bibfnamefont {G.}~\bibnamefont {Xiao}}, \bibinfo
  {author} {\bibfnamefont {K.}~\bibnamefont {Man}}, \bibinfo {author}
  {\bibfnamefont {J.}~\bibnamefont {Dang}}, \bibinfo {author} {\bibfnamefont
  {X.}~\bibnamefont {Cai}}, \bibinfo {author} {\bibfnamefont {Y.}~\bibnamefont
  {Wang}}, \bibinfo {author} {\bibfnamefont {J.}~\bibnamefont {Tang}}, \bibinfo
  {author} {\bibfnamefont {W.}~\bibnamefont {Qiao}}, \bibinfo {author}
  {\bibfnamefont {Y.}~\bibnamefont {Rao}}, \bibinfo {author} {\bibfnamefont
  {Y.}~\bibnamefont {He}}, \bibinfo {author} {\bibfnamefont {L.}~\bibnamefont
  {Mao}}, \ and\ \bibinfo {author} {\bibfnamefont {Z.}~\bibnamefont {Zhou}},\
  }\href {\doibase https://doi.org/10.1016/S0168-9002(02)00475-8} {\bibfield
  {journal} {\bibinfo  {journal} {Nucl. Instrum. Methods Phys. Res., Sect. A}\
  }\textbf {\bibinfo {volume} {488}},\ \bibinfo {pages} {11} (\bibinfo {year}
  {2002})}\BibitemShut {NoStop}%
\bibitem [{\citenamefont {Zhan}\ \emph {et~al.}(2010)\citenamefont {Zhan},
  \citenamefont {Xu}, \citenamefont {Xiao}, \citenamefont {Xia}, \citenamefont
  {Zhao},\ and\ \citenamefont {Yuan}}]{ZHAN2010694c}%
  \BibitemOpen
  \bibfield  {author} {\bibinfo {author} {\bibfnamefont {W.}~\bibnamefont
  {Zhan}}, \bibinfo {author} {\bibfnamefont {H.}~\bibnamefont {Xu}}, \bibinfo
  {author} {\bibfnamefont {G.}~\bibnamefont {Xiao}}, \bibinfo {author}
  {\bibfnamefont {J.}~\bibnamefont {Xia}}, \bibinfo {author} {\bibfnamefont
  {H.}~\bibnamefont {Zhao}}, \ and\ \bibinfo {author} {\bibfnamefont
  {Y.}~\bibnamefont {Yuan}},\ }\href {\doibase
  https://doi.org/10.1016/j.nuclphysa.2010.01.126} {\bibfield  {journal}
  {\bibinfo  {journal} {Nucl. Phys. A}\ }\textbf {\bibinfo {volume} {834}},\
  \bibinfo {pages} {694c} (\bibinfo {year} {2010})}\BibitemShut {NoStop}%
\bibitem [{\citenamefont {Sun}\ \emph {et~al.}(2018)\citenamefont {Sun},
  \citenamefont {Zhao}, \citenamefont {Zhang}, \citenamefont {Sheng},
  \citenamefont {Sun}, \citenamefont {Tanihata}, \citenamefont {Terashima},
  \citenamefont {Zheng}, \citenamefont {Zhu}, \citenamefont {Duan},
  \citenamefont {He}, \citenamefont {Hu}, \citenamefont {Li}, \citenamefont
  {Lin}, \citenamefont {Lin}, \citenamefont {Liu}, \citenamefont {Liu},
  \citenamefont {Lu}, \citenamefont {Ma}, \citenamefont {Mao}, \citenamefont
  {Tian}, \citenamefont {Wang}, \citenamefont {Wang}, \citenamefont {Wang},
  \citenamefont {Xia}, \citenamefont {Xu}, \citenamefont {Xu}, \citenamefont
  {Xu}, \citenamefont {Yang}, \citenamefont {Yin}, \citenamefont {Yuan},
  \citenamefont {Zhan}, \citenamefont {Zhang},\ and\ \citenamefont
  {Zhou}}]{SUN201878}%
  \BibitemOpen
  \bibfield  {author} {\bibinfo {author} {\bibfnamefont {B.-H.}\ \bibnamefont
  {Sun}}, \bibinfo {author} {\bibfnamefont {J.-W.}\ \bibnamefont {Zhao}},
  \bibinfo {author} {\bibfnamefont {X.-H.}\ \bibnamefont {Zhang}}, \bibinfo
  {author} {\bibfnamefont {L.-N.}\ \bibnamefont {Sheng}}, \bibinfo {author}
  {\bibfnamefont {Z.-Y.}\ \bibnamefont {Sun}}, \bibinfo {author} {\bibfnamefont
  {I.}~\bibnamefont {Tanihata}}, \bibinfo {author} {\bibfnamefont
  {S.}~\bibnamefont {Terashima}}, \bibinfo {author} {\bibfnamefont
  {Y.}~\bibnamefont {Zheng}}, \bibinfo {author} {\bibfnamefont {L.-H.}\
  \bibnamefont {Zhu}}, \bibinfo {author} {\bibfnamefont {L.-M.}\ \bibnamefont
  {Duan}}, \bibinfo {author} {\bibfnamefont {L.-C.}\ \bibnamefont {He}},
  \bibinfo {author} {\bibfnamefont {R.-J.}\ \bibnamefont {Hu}}, \bibinfo
  {author} {\bibfnamefont {G.-S.}\ \bibnamefont {Li}}, \bibinfo {author}
  {\bibfnamefont {W.-J.}\ \bibnamefont {Lin}}, \bibinfo {author} {\bibfnamefont
  {W.-P.}\ \bibnamefont {Lin}}, \bibinfo {author} {\bibfnamefont {C.-Y.}\
  \bibnamefont {Liu}}, \bibinfo {author} {\bibfnamefont {Z.}~\bibnamefont
  {Liu}}, \bibinfo {author} {\bibfnamefont {C.-G.}\ \bibnamefont {Lu}},
  \bibinfo {author} {\bibfnamefont {X.-W.}\ \bibnamefont {Ma}}, \bibinfo
  {author} {\bibfnamefont {L.-J.}\ \bibnamefont {Mao}}, \bibinfo {author}
  {\bibfnamefont {Y.}~\bibnamefont {Tian}}, \bibinfo {author} {\bibfnamefont
  {F.}~\bibnamefont {Wang}}, \bibinfo {author} {\bibfnamefont {M.}~\bibnamefont
  {Wang}}, \bibinfo {author} {\bibfnamefont {S.-T.}\ \bibnamefont {Wang}},
  \bibinfo {author} {\bibfnamefont {J.-W.}\ \bibnamefont {Xia}}, \bibinfo
  {author} {\bibfnamefont {X.-D.}\ \bibnamefont {Xu}}, \bibinfo {author}
  {\bibfnamefont {H.-S.}\ \bibnamefont {Xu}}, \bibinfo {author} {\bibfnamefont
  {Z.-G.}\ \bibnamefont {Xu}}, \bibinfo {author} {\bibfnamefont {J.-C.}\
  \bibnamefont {Yang}}, \bibinfo {author} {\bibfnamefont {D.-Y.}\ \bibnamefont
  {Yin}}, \bibinfo {author} {\bibfnamefont {Y.-J.}\ \bibnamefont {Yuan}},
  \bibinfo {author} {\bibfnamefont {W.-L.}\ \bibnamefont {Zhan}}, \bibinfo
  {author} {\bibfnamefont {Y.-H.}\ \bibnamefont {Zhang}}, \ and\ \bibinfo
  {author} {\bibfnamefont {X.-H.}\ \bibnamefont {Zhou}},\ }\href {\doibase
  https://doi.org/10.1016/j.scib.2017.12.005} {\bibfield  {journal} {\bibinfo
  {journal} {Sci. Bull.}\ }\textbf {\bibinfo {volume} {63}},\ \bibinfo {pages}
  {78} (\bibinfo {year} {2018})}\BibitemShut {NoStop}%
\bibitem [{\citenamefont {Lin}\ \emph {et~al.}(2017)\citenamefont {Lin},
  \citenamefont {Zhao}, \citenamefont {Sun}, \citenamefont {He}, \citenamefont
  {Lin}, \citenamefont {Liu}, \citenamefont {Tanihata}, \citenamefont
  {Terashima}, \citenamefont {Tian}, \citenamefont {Wang}, \citenamefont
  {Wang}, \citenamefont {Zhang}, \citenamefont {Zhang}, \citenamefont {Zhu},
  \citenamefont {Duan}, \citenamefont {Hu}, \citenamefont {Liu}, \citenamefont
  {Lu}, \citenamefont {Ren}, \citenamefont {Sheng}, \citenamefont {Sun},
  \citenamefont {Wang}, \citenamefont {Wang}, \citenamefont {Xu},\ and\
  \citenamefont {Zheng}}]{Lin_2017}%
  \BibitemOpen
  \bibfield  {author} {\bibinfo {author} {\bibfnamefont {W.-J.}\ \bibnamefont
  {Lin}}, \bibinfo {author} {\bibfnamefont {J.-W.}\ \bibnamefont {Zhao}},
  \bibinfo {author} {\bibfnamefont {B.-H.}\ \bibnamefont {Sun}}, \bibinfo
  {author} {\bibfnamefont {L.-C.}\ \bibnamefont {He}}, \bibinfo {author}
  {\bibfnamefont {W.-P.}\ \bibnamefont {Lin}}, \bibinfo {author} {\bibfnamefont
  {C.-Y.}\ \bibnamefont {Liu}}, \bibinfo {author} {\bibfnamefont
  {I.}~\bibnamefont {Tanihata}}, \bibinfo {author} {\bibfnamefont
  {S.}~\bibnamefont {Terashima}}, \bibinfo {author} {\bibfnamefont
  {Y.}~\bibnamefont {Tian}}, \bibinfo {author} {\bibfnamefont {F.}~\bibnamefont
  {Wang}}, \bibinfo {author} {\bibfnamefont {M.}~\bibnamefont {Wang}}, \bibinfo
  {author} {\bibfnamefont {G.-X.}\ \bibnamefont {Zhang}}, \bibinfo {author}
  {\bibfnamefont {X.-H.}\ \bibnamefont {Zhang}}, \bibinfo {author}
  {\bibfnamefont {L.-H.}\ \bibnamefont {Zhu}}, \bibinfo {author} {\bibfnamefont
  {L.-M.}\ \bibnamefont {Duan}}, \bibinfo {author} {\bibfnamefont {R.-J.}\
  \bibnamefont {Hu}}, \bibinfo {author} {\bibfnamefont {Z.}~\bibnamefont
  {Liu}}, \bibinfo {author} {\bibfnamefont {C.-G.}\ \bibnamefont {Lu}},
  \bibinfo {author} {\bibfnamefont {P.-P.}\ \bibnamefont {Ren}}, \bibinfo
  {author} {\bibfnamefont {L.-N.}\ \bibnamefont {Sheng}}, \bibinfo {author}
  {\bibfnamefont {Z.-Y.}\ \bibnamefont {Sun}}, \bibinfo {author} {\bibfnamefont
  {S.-T.}\ \bibnamefont {Wang}}, \bibinfo {author} {\bibfnamefont {T.-F.}\
  \bibnamefont {Wang}}, \bibinfo {author} {\bibfnamefont {Z.-G.}\ \bibnamefont
  {Xu}}, \ and\ \bibinfo {author} {\bibfnamefont {Y.}~\bibnamefont {Zheng}},\
  }\href {\doibase 10.1088/1674-1137/41/6/066001} {\bibfield  {journal}
  {\bibinfo  {journal} {Chin. Phys. C}\ }\textbf {\bibinfo {volume} {41}},\
  \bibinfo {pages} {066001} (\bibinfo {year} {2017})}\BibitemShut {NoStop}%
\bibitem [{\citenamefont {Zhao}\ \emph {et~al.}(2019)\citenamefont {Zhao},
  \citenamefont {Sun}, \citenamefont {He}, \citenamefont {Li}, \citenamefont
  {Lin}, \citenamefont {Liu}, \citenamefont {Liu}, \citenamefont {Lu},
  \citenamefont {Shen}, \citenamefont {Sun}, \citenamefont {Sun}, \citenamefont
  {Tanihata}, \citenamefont {Terashima}, \citenamefont {Tran}, \citenamefont
  {Wang}, \citenamefont {Wang}, \citenamefont {Wang}, \citenamefont {Wei},
  \citenamefont {Xu}, \citenamefont {Zhu}, \citenamefont {Zhang}, \citenamefont
  {Zhang}, \citenamefont {Zhang},\ and\ \citenamefont {Zhou}}]{ZHAO201995}%
  \BibitemOpen
  \bibfield  {author} {\bibinfo {author} {\bibfnamefont {J.}~\bibnamefont
  {Zhao}}, \bibinfo {author} {\bibfnamefont {B.}~\bibnamefont {Sun}}, \bibinfo
  {author} {\bibfnamefont {L.}~\bibnamefont {He}}, \bibinfo {author}
  {\bibfnamefont {G.}~\bibnamefont {Li}}, \bibinfo {author} {\bibfnamefont
  {W.}~\bibnamefont {Lin}}, \bibinfo {author} {\bibfnamefont {C.}~\bibnamefont
  {Liu}}, \bibinfo {author} {\bibfnamefont {Z.}~\bibnamefont {Liu}}, \bibinfo
  {author} {\bibfnamefont {C.}~\bibnamefont {Lu}}, \bibinfo {author}
  {\bibfnamefont {D.}~\bibnamefont {Shen}}, \bibinfo {author} {\bibfnamefont
  {Y.}~\bibnamefont {Sun}}, \bibinfo {author} {\bibfnamefont {Z.}~\bibnamefont
  {Sun}}, \bibinfo {author} {\bibfnamefont {I.}~\bibnamefont {Tanihata}},
  \bibinfo {author} {\bibfnamefont {S.}~\bibnamefont {Terashima}}, \bibinfo
  {author} {\bibfnamefont {D.}~\bibnamefont {Tran}}, \bibinfo {author}
  {\bibfnamefont {F.}~\bibnamefont {Wang}}, \bibinfo {author} {\bibfnamefont
  {J.}~\bibnamefont {Wang}}, \bibinfo {author} {\bibfnamefont {S.}~\bibnamefont
  {Wang}}, \bibinfo {author} {\bibfnamefont {X.}~\bibnamefont {Wei}}, \bibinfo
  {author} {\bibfnamefont {X.}~\bibnamefont {Xu}}, \bibinfo {author}
  {\bibfnamefont {L.}~\bibnamefont {Zhu}}, \bibinfo {author} {\bibfnamefont
  {J.}~\bibnamefont {Zhang}}, \bibinfo {author} {\bibfnamefont
  {X.}~\bibnamefont {Zhang}}, \bibinfo {author} {\bibfnamefont
  {Y.}~\bibnamefont {Zhang}}, \ and\ \bibinfo {author} {\bibfnamefont
  {Z.}~\bibnamefont {Zhou}},\ }\href {\doibase
  https://doi.org/10.1016/j.nima.2019.03.063} {\bibfield  {journal} {\bibinfo
  {journal} {Nucl. Instrum. Methods Phys. Res., Sect. A}\ }\textbf {\bibinfo
  {volume} {930}},\ \bibinfo {pages} {95} (\bibinfo {year} {2019})}\BibitemShut
  {NoStop}%
\bibitem [{\citenamefont {Zhao}\ \emph {et~al.}(2020)\citenamefont {Zhao},
  \citenamefont {Sun}, \citenamefont {Terashima}, \citenamefont {He},
  \citenamefont {Lu}, \citenamefont {Li}, \citenamefont {Liu}, \citenamefont
  {Sun}, \citenamefont {Tanihata}, \citenamefont {Wang}, \citenamefont {Wang},
  \citenamefont {Wang}, \citenamefont {Wei}, \citenamefont {Zhang},
  \citenamefont {Zhu},\ and\ \citenamefont {Zhang}}]{Zhao:2020seq}%
  \BibitemOpen
  \bibfield  {author} {\bibinfo {author} {\bibfnamefont {J.}~\bibnamefont
  {Zhao}}, \bibinfo {author} {\bibfnamefont {B.}~\bibnamefont {Sun}}, \bibinfo
  {author} {\bibfnamefont {S.}~\bibnamefont {Terashima}}, \bibinfo {author}
  {\bibfnamefont {L.}~\bibnamefont {He}}, \bibinfo {author} {\bibfnamefont
  {C.}~\bibnamefont {Lu}}, \bibinfo {author} {\bibfnamefont {G.}~\bibnamefont
  {Li}}, \bibinfo {author} {\bibfnamefont {Z.}~\bibnamefont {Liu}}, \bibinfo
  {author} {\bibfnamefont {Z.}~\bibnamefont {Sun}}, \bibinfo {author}
  {\bibfnamefont {I.}~\bibnamefont {Tanihata}}, \bibinfo {author}
  {\bibfnamefont {F.}~\bibnamefont {Wang}}, \bibinfo {author} {\bibfnamefont
  {M.}~\bibnamefont {Wang}}, \bibinfo {author} {\bibfnamefont {S.}~\bibnamefont
  {Wang}}, \bibinfo {author} {\bibfnamefont {X.}~\bibnamefont {Wei}}, \bibinfo
  {author} {\bibfnamefont {J.}~\bibnamefont {Zhang}}, \bibinfo {author}
  {\bibfnamefont {L.}~\bibnamefont {Zhu}}, \ and\ \bibinfo {author}
  {\bibfnamefont {X.}~\bibnamefont {Zhang}},\ }\href {\doibase
  10.7566/JPSCP.32.010023} {\bibfield  {journal} {\bibinfo  {journal} {JPS
  Conf. Proc.}\ }\textbf {\bibinfo {volume} {32}},\ \bibinfo {pages} {010023}
  (\bibinfo {year} {2020})}\BibitemShut {NoStop}%
\bibitem [{\citenamefont {Zhang}\ \emph {et~al.}(2015)\citenamefont {Zhang},
  \citenamefont {Tang}, \citenamefont {Ma}, \citenamefont {Lu}, \citenamefont
  {Yang}, \citenamefont {Wang}, \citenamefont {Yu}, \citenamefont {Yue},
  \citenamefont {Fang}, \citenamefont {Yan}, \citenamefont {Zhou},
  \citenamefont {Wang}, \citenamefont {Sun}, \citenamefont {Sun}, \citenamefont
  {Duan},\ and\ \citenamefont {Sun}}]{ZHANG2015389}%
  \BibitemOpen
  \bibfield  {author} {\bibinfo {author} {\bibfnamefont {X.}~\bibnamefont
  {Zhang}}, \bibinfo {author} {\bibfnamefont {S.}~\bibnamefont {Tang}},
  \bibinfo {author} {\bibfnamefont {P.}~\bibnamefont {Ma}}, \bibinfo {author}
  {\bibfnamefont {C.}~\bibnamefont {Lu}}, \bibinfo {author} {\bibfnamefont
  {H.}~\bibnamefont {Yang}}, \bibinfo {author} {\bibfnamefont {S.}~\bibnamefont
  {Wang}}, \bibinfo {author} {\bibfnamefont {Y.}~\bibnamefont {Yu}}, \bibinfo
  {author} {\bibfnamefont {K.}~\bibnamefont {Yue}}, \bibinfo {author}
  {\bibfnamefont {F.}~\bibnamefont {Fang}}, \bibinfo {author} {\bibfnamefont
  {D.}~\bibnamefont {Yan}}, \bibinfo {author} {\bibfnamefont {Y.}~\bibnamefont
  {Zhou}}, \bibinfo {author} {\bibfnamefont {Z.}~\bibnamefont {Wang}}, \bibinfo
  {author} {\bibfnamefont {Y.}~\bibnamefont {Sun}}, \bibinfo {author}
  {\bibfnamefont {Z.}~\bibnamefont {Sun}}, \bibinfo {author} {\bibfnamefont
  {L.}~\bibnamefont {Duan}}, \ and\ \bibinfo {author} {\bibfnamefont
  {B.}~\bibnamefont {Sun}},\ }\href {\doibase
  https://doi.org/10.1016/j.nima.2015.06.022} {\bibfield  {journal} {\bibinfo
  {journal} {Nucl. Instrum. Methods Phys. Res., Sect. A}\ }\textbf {\bibinfo
  {volume} {795}},\ \bibinfo {pages} {389} (\bibinfo {year}
  {2015})}\BibitemShut {NoStop}%
\bibitem [{\citenamefont {Sun}\ \emph {et~al.}(2019)\citenamefont {Sun},
  \citenamefont {Sun}, \citenamefont {Wang}, \citenamefont {Zhang},
  \citenamefont {Sun}, \citenamefont {Yan}, \citenamefont {Tang}, \citenamefont
  {Yu}, \citenamefont {Yue}, \citenamefont {Duan}, \citenamefont {Yang},
  \citenamefont {Lu}, \citenamefont {Fang}, \citenamefont {Ma},\ and\
  \citenamefont {Su}}]{SUN2019390}%
  \BibitemOpen
  \bibfield  {author} {\bibinfo {author} {\bibfnamefont {Y.}~\bibnamefont
  {Sun}}, \bibinfo {author} {\bibfnamefont {Z.}~\bibnamefont {Sun}}, \bibinfo
  {author} {\bibfnamefont {S.}~\bibnamefont {Wang}}, \bibinfo {author}
  {\bibfnamefont {X.}~\bibnamefont {Zhang}}, \bibinfo {author} {\bibfnamefont
  {Y.}~\bibnamefont {Sun}}, \bibinfo {author} {\bibfnamefont {D.}~\bibnamefont
  {Yan}}, \bibinfo {author} {\bibfnamefont {S.}~\bibnamefont {Tang}}, \bibinfo
  {author} {\bibfnamefont {Y.}~\bibnamefont {Yu}}, \bibinfo {author}
  {\bibfnamefont {K.}~\bibnamefont {Yue}}, \bibinfo {author} {\bibfnamefont
  {L.}~\bibnamefont {Duan}}, \bibinfo {author} {\bibfnamefont {H.}~\bibnamefont
  {Yang}}, \bibinfo {author} {\bibfnamefont {C.}~\bibnamefont {Lu}}, \bibinfo
  {author} {\bibfnamefont {F.}~\bibnamefont {Fang}}, \bibinfo {author}
  {\bibfnamefont {P.}~\bibnamefont {Ma}}, \ and\ \bibinfo {author}
  {\bibfnamefont {H.}~\bibnamefont {Su}},\ }\href {\doibase
  https://doi.org/10.1016/j.nima.2019.02.067} {\bibfield  {journal} {\bibinfo
  {journal} {Nucl. Instrum. Methods Phys. Res., Sect. A}\ }\textbf {\bibinfo
  {volume} {927}},\ \bibinfo {pages} {390} (\bibinfo {year}
  {2019})}\BibitemShut {NoStop}%
\bibitem [{\citenamefont {Sun}(2020)}]{BaoHS2020CSB}%
  \BibitemOpen
  \bibfield  {author} {\bibinfo {author} {\bibfnamefont {B.~H.}\ \bibnamefont
  {Sun}},\ }\href {\doibase 10.1360/TB-2020-0906} {\bibfield  {journal}
  {\bibinfo  {journal} {Chin. Sci. Bull.}\ }\textbf {\bibinfo {volume} {65}},\
  \bibinfo {pages} {3886} (\bibinfo {year} {2020})}\BibitemShut {NoStop}%
\bibitem [{\citenamefont {Zeitlin}\ \emph {et~al.}(2001)\citenamefont
  {Zeitlin}, \citenamefont {Fukumura}, \citenamefont {Heilbronn}, \citenamefont
  {Iwata}, \citenamefont {Miller},\ and\ \citenamefont
  {Murakami}}]{PhysRevC.64.024902}%
  \BibitemOpen
  \bibfield  {author} {\bibinfo {author} {\bibfnamefont {C.}~\bibnamefont
  {Zeitlin}}, \bibinfo {author} {\bibfnamefont {A.}~\bibnamefont {Fukumura}},
  \bibinfo {author} {\bibfnamefont {L.}~\bibnamefont {Heilbronn}}, \bibinfo
  {author} {\bibfnamefont {Y.}~\bibnamefont {Iwata}}, \bibinfo {author}
  {\bibfnamefont {J.}~\bibnamefont {Miller}}, \ and\ \bibinfo {author}
  {\bibfnamefont {T.}~\bibnamefont {Murakami}},\ }\href {\doibase
  10.1103/PhysRevC.64.024902} {\bibfield  {journal} {\bibinfo  {journal} {Phys.
  Rev. C}\ }\textbf {\bibinfo {volume} {64}},\ \bibinfo {pages} {024902}
  (\bibinfo {year} {2001})}\BibitemShut {NoStop}%
\bibitem [{\citenamefont {Webber}\ \emph
  {et~al.}(1990{\natexlab{b}})\citenamefont {Webber}, \citenamefont {Kish},\
  and\ \citenamefont {Schrier}}]{PhysRevC.41.520}%
  \BibitemOpen
  \bibfield  {author} {\bibinfo {author} {\bibfnamefont {W.~R.}\ \bibnamefont
  {Webber}}, \bibinfo {author} {\bibfnamefont {J.~C.}\ \bibnamefont {Kish}}, \
  and\ \bibinfo {author} {\bibfnamefont {D.~A.}\ \bibnamefont {Schrier}},\
  }\href {\doibase 10.1103/PhysRevC.41.520} {\bibfield  {journal} {\bibinfo
  {journal} {Phys. Rev. C}\ }\textbf {\bibinfo {volume} {41}},\ \bibinfo
  {pages} {520} (\bibinfo {year} {1990}{\natexlab{b}})}\BibitemShut {NoStop}%
\bibitem [{\citenamefont {Yamaguchi}\ \emph {et~al.}(2010)\citenamefont
  {Yamaguchi}, \citenamefont {Fukuda}, \citenamefont {Fukuda}, \citenamefont
  {Fan}, \citenamefont {Hachiuma}, \citenamefont {Kanazawa}, \citenamefont
  {Kitagawa}, \citenamefont {Kuboki}, \citenamefont {Lantz}, \citenamefont
  {Mihara}, \citenamefont {Nagashima}, \citenamefont {Namihira}, \citenamefont
  {Nishimura}, \citenamefont {Okuma}, \citenamefont {Ohtsubo}, \citenamefont
  {Sato}, \citenamefont {Suzuki}, \citenamefont {Takechi},\ and\ \citenamefont
  {Xu}}]{PhysRevC.82.014609}%
  \BibitemOpen
  \bibfield  {author} {\bibinfo {author} {\bibfnamefont {T.}~\bibnamefont
  {Yamaguchi}}, \bibinfo {author} {\bibfnamefont {M.}~\bibnamefont {Fukuda}},
  \bibinfo {author} {\bibfnamefont {S.}~\bibnamefont {Fukuda}}, \bibinfo
  {author} {\bibfnamefont {G.~W.}\ \bibnamefont {Fan}}, \bibinfo {author}
  {\bibfnamefont {I.}~\bibnamefont {Hachiuma}}, \bibinfo {author}
  {\bibfnamefont {M.}~\bibnamefont {Kanazawa}}, \bibinfo {author}
  {\bibfnamefont {A.}~\bibnamefont {Kitagawa}}, \bibinfo {author}
  {\bibfnamefont {T.}~\bibnamefont {Kuboki}}, \bibinfo {author} {\bibfnamefont
  {M.}~\bibnamefont {Lantz}}, \bibinfo {author} {\bibfnamefont
  {M.}~\bibnamefont {Mihara}}, \bibinfo {author} {\bibfnamefont
  {M.}~\bibnamefont {Nagashima}}, \bibinfo {author} {\bibfnamefont
  {K.}~\bibnamefont {Namihira}}, \bibinfo {author} {\bibfnamefont
  {D.}~\bibnamefont {Nishimura}}, \bibinfo {author} {\bibfnamefont
  {Y.}~\bibnamefont {Okuma}}, \bibinfo {author} {\bibfnamefont
  {T.}~\bibnamefont {Ohtsubo}}, \bibinfo {author} {\bibfnamefont
  {S.}~\bibnamefont {Sato}}, \bibinfo {author} {\bibfnamefont {T.}~\bibnamefont
  {Suzuki}}, \bibinfo {author} {\bibfnamefont {M.}~\bibnamefont {Takechi}}, \
  and\ \bibinfo {author} {\bibfnamefont {W.}~\bibnamefont {Xu}},\ }\href
  {\doibase 10.1103/PhysRevC.82.014609} {\bibfield  {journal} {\bibinfo
  {journal} {Phys. Rev. C}\ }\textbf {\bibinfo {volume} {82}},\ \bibinfo
  {pages} {014609} (\bibinfo {year} {2010})}\BibitemShut {NoStop}%
\bibitem [{\citenamefont {Iancu}\ \emph {et~al.}(2005)\citenamefont {Iancu},
  \citenamefont {Flesch},\ and\ \citenamefont {Heinrich}}]{IANCU2005525}%
  \BibitemOpen
  \bibfield  {author} {\bibinfo {author} {\bibfnamefont {G.}~\bibnamefont
  {Iancu}}, \bibinfo {author} {\bibfnamefont {F.}~\bibnamefont {Flesch}}, \
  and\ \bibinfo {author} {\bibfnamefont {W.}~\bibnamefont {Heinrich}},\ }\href
  {\doibase https://doi.org/10.1016/j.radmeas.2004.10.011} {\bibfield
  {journal} {\bibinfo  {journal} {Radiat. Meas.}\ }\textbf {\bibinfo {volume}
  {39}},\ \bibinfo {pages} {525} (\bibinfo {year} {2005})}\BibitemShut
  {NoStop}%
\bibitem [{\citenamefont {Zeitlin}\ \emph {et~al.}(2008)\citenamefont
  {Zeitlin}, \citenamefont {Guetersloh}, \citenamefont {Heilbronn},
  \citenamefont {Miller}, \citenamefont {Fukumura}, \citenamefont {Iwata},
  \citenamefont {Murakami}, \citenamefont {Sihver},\ and\ \citenamefont
  {Mancusi}}]{PhysRevC.77.034605}%
  \BibitemOpen
  \bibfield  {author} {\bibinfo {author} {\bibfnamefont {C.}~\bibnamefont
  {Zeitlin}}, \bibinfo {author} {\bibfnamefont {S.}~\bibnamefont {Guetersloh}},
  \bibinfo {author} {\bibfnamefont {L.}~\bibnamefont {Heilbronn}}, \bibinfo
  {author} {\bibfnamefont {J.}~\bibnamefont {Miller}}, \bibinfo {author}
  {\bibfnamefont {A.}~\bibnamefont {Fukumura}}, \bibinfo {author}
  {\bibfnamefont {Y.}~\bibnamefont {Iwata}}, \bibinfo {author} {\bibfnamefont
  {T.}~\bibnamefont {Murakami}}, \bibinfo {author} {\bibfnamefont
  {L.}~\bibnamefont {Sihver}}, \ and\ \bibinfo {author} {\bibfnamefont
  {D.}~\bibnamefont {Mancusi}},\ }\href {\doibase 10.1103/PhysRevC.77.034605}
  {\bibfield  {journal} {\bibinfo  {journal} {Phys. Rev. C}\ }\textbf {\bibinfo
  {volume} {77}},\ \bibinfo {pages} {034605} (\bibinfo {year}
  {2008})}\BibitemShut {NoStop}%
\bibitem [{\citenamefont {Zeitlin}\ \emph {et~al.}(1997)\citenamefont
  {Zeitlin}, \citenamefont {Heilbronn}, \citenamefont {Miller}, \citenamefont
  {Rademacher}, \citenamefont {Borak}, \citenamefont {Carter}, \citenamefont
  {Frankel}, \citenamefont {Schimmerling},\ and\ \citenamefont
  {Stronach}}]{PhysRevC.56.388}%
  \BibitemOpen
  \bibfield  {author} {\bibinfo {author} {\bibfnamefont {C.}~\bibnamefont
  {Zeitlin}}, \bibinfo {author} {\bibfnamefont {L.}~\bibnamefont {Heilbronn}},
  \bibinfo {author} {\bibfnamefont {J.}~\bibnamefont {Miller}}, \bibinfo
  {author} {\bibfnamefont {S.~E.}\ \bibnamefont {Rademacher}}, \bibinfo
  {author} {\bibfnamefont {T.}~\bibnamefont {Borak}}, \bibinfo {author}
  {\bibfnamefont {T.~R.}\ \bibnamefont {Carter}}, \bibinfo {author}
  {\bibfnamefont {K.~A.}\ \bibnamefont {Frankel}}, \bibinfo {author}
  {\bibfnamefont {W.}~\bibnamefont {Schimmerling}}, \ and\ \bibinfo {author}
  {\bibfnamefont {C.~E.}\ \bibnamefont {Stronach}},\ }\href {\doibase
  10.1103/PhysRevC.56.388} {\bibfield  {journal} {\bibinfo  {journal} {Phys.
  Rev. C}\ }\textbf {\bibinfo {volume} {56}},\ \bibinfo {pages} {388} (\bibinfo
  {year} {1997})}\BibitemShut {NoStop}%
\bibitem [{\citenamefont {Ricciardi}\ \emph {et~al.}(2004)\citenamefont
  {Ricciardi}, \citenamefont {Ignatyuk}, \citenamefont {Kelić}, \citenamefont
  {Napolitani}, \citenamefont {Rejmund}, \citenamefont {Schmidt},\ and\
  \citenamefont {Yordanov}}]{RICCIARDI2004299}%
  \BibitemOpen
  \bibfield  {author} {\bibinfo {author} {\bibfnamefont {M.}~\bibnamefont
  {Ricciardi}}, \bibinfo {author} {\bibfnamefont {A.}~\bibnamefont {Ignatyuk}},
  \bibinfo {author} {\bibfnamefont {A.}~\bibnamefont {Kelić}}, \bibinfo
  {author} {\bibfnamefont {P.}~\bibnamefont {Napolitani}}, \bibinfo {author}
  {\bibfnamefont {F.}~\bibnamefont {Rejmund}}, \bibinfo {author} {\bibfnamefont
  {K.-H.}\ \bibnamefont {Schmidt}}, \ and\ \bibinfo {author} {\bibfnamefont
  {O.}~\bibnamefont {Yordanov}},\ }\href {\doibase
  https://doi.org/10.1016/j.nuclphysa.2004.01.069} {\bibfield  {journal}
  {\bibinfo  {journal} {Nucl. Phys. A}\ }\textbf {\bibinfo {volume} {733}},\
  \bibinfo {pages} {299} (\bibinfo {year} {2004})}\BibitemShut {NoStop}%
\bibitem [{\citenamefont {Mei}\ \emph {et~al.}(2018)\citenamefont {Mei},
  \citenamefont {Tu},\ and\ \citenamefont {Wang}}]{PhysRevC.97.044619}%
  \BibitemOpen
  \bibfield  {author} {\bibinfo {author} {\bibfnamefont {B.}~\bibnamefont
  {Mei}}, \bibinfo {author} {\bibfnamefont {X.~L.}\ \bibnamefont {Tu}}, \ and\
  \bibinfo {author} {\bibfnamefont {M.}~\bibnamefont {Wang}},\ }\href {\doibase
  10.1103/PhysRevC.97.044619} {\bibfield  {journal} {\bibinfo  {journal} {Phys.
  Rev. C}\ }\textbf {\bibinfo {volume} {97}},\ \bibinfo {pages} {044619}
  (\bibinfo {year} {2018})}\BibitemShut {NoStop}%
\bibitem [{\citenamefont {Mei}\ \emph {et~al.}(2022)\citenamefont {Mei},
  \citenamefont {Tu}, \citenamefont {Zhang}, \citenamefont {Wang},
  \citenamefont {Guan}, \citenamefont {Mai}, \citenamefont {Zeng},
  \citenamefont {Tu}, \citenamefont {Sun}, \citenamefont {Tang}, \citenamefont
  {Yu}, \citenamefont {Fang}, \citenamefont {Yan}, \citenamefont {Jin},
  \citenamefont {Zhao}, \citenamefont {Ma},\ and\ \citenamefont
  {Zhang}}]{PhysRevC.105.064604}%
  \BibitemOpen
  \bibfield  {author} {\bibinfo {author} {\bibfnamefont {B.}~\bibnamefont
  {Mei}}, \bibinfo {author} {\bibfnamefont {J.}~\bibnamefont {Tu}}, \bibinfo
  {author} {\bibfnamefont {X.}~\bibnamefont {Zhang}}, \bibinfo {author}
  {\bibfnamefont {S.}~\bibnamefont {Wang}}, \bibinfo {author} {\bibfnamefont
  {Y.}~\bibnamefont {Guan}}, \bibinfo {author} {\bibfnamefont {Z.}~\bibnamefont
  {Mai}}, \bibinfo {author} {\bibfnamefont {N.}~\bibnamefont {Zeng}}, \bibinfo
  {author} {\bibfnamefont {X.}~\bibnamefont {Tu}}, \bibinfo {author}
  {\bibfnamefont {Z.}~\bibnamefont {Sun}}, \bibinfo {author} {\bibfnamefont
  {S.}~\bibnamefont {Tang}}, \bibinfo {author} {\bibfnamefont {Y.}~\bibnamefont
  {Yu}}, \bibinfo {author} {\bibfnamefont {F.}~\bibnamefont {Fang}}, \bibinfo
  {author} {\bibfnamefont {D.}~\bibnamefont {Yan}}, \bibinfo {author}
  {\bibfnamefont {S.}~\bibnamefont {Jin}}, \bibinfo {author} {\bibfnamefont
  {Y.}~\bibnamefont {Zhao}}, \bibinfo {author} {\bibfnamefont {S.}~\bibnamefont
  {Ma}}, \ and\ \bibinfo {author} {\bibfnamefont {Y.}~\bibnamefont {Zhang}},\
  }\href {\doibase 10.1103/PhysRevC.105.064604} {\bibfield  {journal} {\bibinfo
   {journal} {Phys. Rev. C}\ }\textbf {\bibinfo {volume} {105}},\ \bibinfo
  {pages} {064604} (\bibinfo {year} {2022})}\BibitemShut {NoStop}%
\bibitem [{\citenamefont {Golovchenko}\ \emph {et~al.}(2002)\citenamefont
  {Golovchenko}, \citenamefont {Skvar\ifmmode~\check{c}\else \v{c}\fi{}},
  \citenamefont {Yasuda}, \citenamefont {Giacomelli}, \citenamefont
  {Tretyakova}, \citenamefont {Ili\ifmmode~\acute{c}\else \'{c}\fi{}},
  \citenamefont {Bimbot}, \citenamefont {Toulemonde},\ and\ \citenamefont
  {Murakami}}]{PhysRevC.66.014609}%
  \BibitemOpen
  \bibfield  {author} {\bibinfo {author} {\bibfnamefont {A.~N.}\ \bibnamefont
  {Golovchenko}}, \bibinfo {author} {\bibfnamefont {J.}~\bibnamefont
  {Skvar\ifmmode~\check{c}\else \v{c}\fi{}}}, \bibinfo {author} {\bibfnamefont
  {N.}~\bibnamefont {Yasuda}}, \bibinfo {author} {\bibfnamefont
  {M.}~\bibnamefont {Giacomelli}}, \bibinfo {author} {\bibfnamefont {S.~P.}\
  \bibnamefont {Tretyakova}}, \bibinfo {author} {\bibfnamefont
  {R.}~\bibnamefont {Ili\ifmmode~\acute{c}\else \'{c}\fi{}}}, \bibinfo {author}
  {\bibfnamefont {R.}~\bibnamefont {Bimbot}}, \bibinfo {author} {\bibfnamefont
  {M.}~\bibnamefont {Toulemonde}}, \ and\ \bibinfo {author} {\bibfnamefont
  {T.}~\bibnamefont {Murakami}},\ }\href {\doibase 10.1103/PhysRevC.66.014609}
  {\bibfield  {journal} {\bibinfo  {journal} {Phys. Rev. C}\ }\textbf {\bibinfo
  {volume} {66}},\ \bibinfo {pages} {014609} (\bibinfo {year}
  {2002})}\BibitemShut {NoStop}%
\bibitem [{\citenamefont {Horiuchi}\ \emph {et~al.}(2014)\citenamefont
  {Horiuchi}, \citenamefont {Suzuki},\ and\ \citenamefont
  {Inakura}}]{PhysRevC.89.011601}%
  \BibitemOpen
  \bibfield  {author} {\bibinfo {author} {\bibfnamefont {W.}~\bibnamefont
  {Horiuchi}}, \bibinfo {author} {\bibfnamefont {Y.}~\bibnamefont {Suzuki}}, \
  and\ \bibinfo {author} {\bibfnamefont {T.}~\bibnamefont {Inakura}},\ }\href
  {\doibase 10.1103/PhysRevC.89.011601} {\bibfield  {journal} {\bibinfo
  {journal} {Phys. Rev. C}\ }\textbf {\bibinfo {volume} {89}},\ \bibinfo
  {pages} {011601(R)} (\bibinfo {year} {2014})}\BibitemShut {NoStop}%
\bibitem [{\citenamefont {Bazin}\ \emph {et~al.}(2002)\citenamefont {Bazin},
  \citenamefont {Tarasov}, \citenamefont {Lewitowicz},\ and\ \citenamefont
  {Sorlin}}]{BAZIN2002307}%
  \BibitemOpen
  \bibfield  {author} {\bibinfo {author} {\bibfnamefont {D.}~\bibnamefont
  {Bazin}}, \bibinfo {author} {\bibfnamefont {O.}~\bibnamefont {Tarasov}},
  \bibinfo {author} {\bibfnamefont {M.}~\bibnamefont {Lewitowicz}}, \ and\
  \bibinfo {author} {\bibfnamefont {O.}~\bibnamefont {Sorlin}},\ }\href
  {\doibase https://doi.org/10.1016/S0168-9002(01)01504-2} {\bibfield
  {journal} {\bibinfo  {journal} {Nucl. Instrum. Methods Phys. Res., Sect. A}\
  }\textbf {\bibinfo {volume} {482}},\ \bibinfo {pages} {307} (\bibinfo {year}
  {2002})}\BibitemShut {NoStop}%
\end{thebibliography}
%
 
\end{document}